# Topological Ring-Cavity Laser Formed by Honeycomb Photonic Crystals

Xiao-Chen Sun and Xiao Hu*
International Center for Materials Nanoarchitectonics (WPI-MANA)
National Institute for Materials Science (NIMS), Tsukuba 305-0044, Japan
**e-mail: HU.Xiao(at)nims.go.jp*

*Abstract*

We clarify theoretically that the topological ring-cavity (TRC) modes propagating along the interface between two honeycomb-type photonic crystals distinct in topology can be exploited for achieving stable single-mode lasing, with the maximal intensity larger than a whispering-gallery-mode counterpart by order of magnitude. Especially, we show that the TRC modes located at the bulk band gap center benefit maximally from the gain profile since they are most concentrated and uniform along the interface, and that, inheriting from the Dirac-like dispersion of topological interface states, they are separated in frequency from each other and from other photonic modes, both favoring intrinsically single-mode lasing. A TRC mode running in a specific direction with desired orbital angular momentum can be stimulated selectively by injecting a circularly polarized beam. With the scheme proposed in the present work one can fabricate a topological laser of tens of micro-meters in linear size by means of advanced semiconductor nanotechnologies, which generates chiral laser beams ideal for novel photonic functions.

## I. INTRODUCTION

Recent years have seen a surge of research interests on topological states of matter [1-9]. The topological edge states are characterized by topological invariants, thus immune to scattering from possible sample imperfections, robust against disorder, and useful for innovative applications. Haldane and Raghu proposed the photonic analog of quantum Hall effect, which was confirmed experimentally in a gyro-magnetic system at room temperature [10-13]. Subsequent efforts have been devoted to explore photonic topological insulators with unidirectional electromagnetic（EM）propagations, including coupled-resonator optical waveguides (CROW) [14, 15], helical waveguides with artificial gauge field [16], metamaterial structures with bi-anisotropy [17], and dielectric photonic crystals (PhCs) [18, 19], as well as a rich wealth of photonic systems with various topological features [20-24]. Releasing the usage of the Faraday effect in achieving topological photonic states, which requires external magnetic field and is effective only below the infrared band, these recent approaches in principle make on-chip integration of topological features into various photonic functions possible, and especially those based on dielectric materials can potentially push the operation frequency of topological photonic states to visible light. Recently, topological lasers have been realized based on chiral edge states with broken time-reversal (TR) symmetry [25], helical edge states in the CROW structure with TR symmetry [26, 27], bulk states in honeycomb-type PhCs [28] and valley edge modes [29], which demonstrated successfully improvement in the slope efficiency and the single-mode feature over the trivial counterparts, and meanwhile revealed a new aspect of non-Hermitian topology.

In this work, we adopt a design where the TR symmetric photonic topology is derived purely from the local and periodic deformations of honeycomb structure in dielectric PhCs [18]. We clarify theoretically how the intrinsic topological features can improve the lasing performance, apart from the robustness against possible deformations and disorder. Inheriting from the Dirac-like dispersion, the topological interface states are separated in frequency from each other and from other topological interface states, which favors single-mode lasing up to high gain values. Originating from the spin-momentum locking in the photonic analog of the quantum spin Hall effect, the lasing ring-cavity modes carry orbital angular momenta (OAM) along the closed interface governed by the pseudospin defined in hexagonal unit cells. On one hand, this feature provides a handle for selective pumping towards lasing, on the other hand, it enables generation of vortex beams useful for advanced optic applications. Because of the simplicity of the design all these features of topology origin can be realized in laser devices as small as tens of micro-meters, which is favorable for applications.

The remaining part of this paper is organized as follows. In Sec. II, we start by providing a detailed formalism which equips ourselves for the following calculations. We then proceed to calculate the eigen helical modes in a topological ring cavity (TRC) formed by a topological PhC surrounded by a trivial PhC. Results obtained by analytical calculations and numerical diagonalizations based on a tight-binding (TB) model, and computer simulations based on the Finite Element Analysis software COMSOL [30] are compared successfully. In order to clarify the lasing phenomenon, we introduce in Sec. III non-linear and non-Hermitian terms into the TB Hamiltonian and investigate its time evolution. Specifically, we demonstrate that the TRC modes at the center of bulk band gap benefit maximally from the gain profile. We unveil that the maximal intensity of the single-mode lasing of the TRC laser is larger than a typical whispering-gallery-mode (WGM) laser by order of magnitude, and that the TRC modes carrying the desired OAM can be driven to lase by injecting circularly polarized light. A discussion is provided in Sec. IV, followed by a summary in Sec. V.

## II. HELICAL RING-CAVITY MODES

### A. Jackiw-Rebbi state and dispersion relation of TRC modes

The present TRC laser is formed by a topological PhC surrounded by a trivial PhC as schematically shown in Fig. 1, with the separation $R$ between neighboring dielectric cylinders inside a hexagonal unit cell and the common lattice constant $a$ satisfying the relation $a/R<3$ and $a/R>3$ respectively ($a/R=3$ corresponds to a honeycomb lattice) [18]. The topological feature of a honeycomb-type PhC can be captured by the TB Hamiltonian with nearest-neighbor (n.n.) couplings [31]

$$\hat{H}=\sum_{\langle n,m\rangle} t_{nm}\hat{c}_n^{\dagger}\hat{c}_m=t_{\rm in}\sum_{\langle n,m\rangle}\hat{c}_n^{\dagger}\hat{c}_m+t_{\rm out}\sum_{\langle n',m'\rangle}\hat{c}_{n'}^{\dagger}\hat{c}_{m'}=\hat{\Psi}^{\dagger}H\hat{\Psi} \tag{1}$$

where $\hat{\Psi}=(\hat{c}_1\quad\cdots\quad\hat{c}_N)^T$, $\hat{c}_n$ $(\hat{c}_n^{\dagger})$ is the photonic annihilation (creation) operator on site $n\in\{0,1,\ldots,N\}$, $t_{nm}(< 0)$ is the coupling between neighboring sites of honeycomb lattice (see Fig. 1): $t_{nm} = t_{\text{in}}$ and $t_{nm} = t_{\text{out}}$ for $\langle n, m\rangle$ inside and between hexagonal unit cells respectively. To be specific, we set $t_{\text{in}} = -2.5t_0$ and $t_{\text{out}} = -5t_0$ for the topological PhC at the device center, $t_{\text{in}} = -15t_0/4$ and $t_{\text{out}} = -15t_0/7$ for the cladding trivial PhC, where $t_0$ $(> 0)$ is the unit of energy. Because the coupling between two cylinders in a PhC decays monotonically with their separation, the above strong-weak texture of the n.n. couplings respecting $C_{6v}$ symmetry corresponds directly to the real-space positioning of the dielectric cylinders (see Fig. 1).

As revealed previously [18], around the band gap opened at Γ point there are two pairs of doubly degenerate states, which can be expressed as $|p_x\rangle$ / $|p_y\rangle$ and $|d_{x^2-y^2}\rangle/|d_{2xy}\rangle$ as shown in Fig. 2(a) where a hexagonal unit cell is taken. We can make linear combinations for them and obtain the following pairs of eigenstates: $|p_\pm\rangle = |p_x\rangle \pm i|p_y\rangle$ and $|d_\pm\rangle = |d_{x^2-y^2}\rangle \pm i|d_{2xy}\rangle$, where the phase changes over the six sites in the unit cell as shown in Fig. 2(b) with constant amplitudes. Going through the six sites counterclockwise, $|p_\pm\rangle$ and $|d_\pm\rangle$ pick up phases of $\pm 2\pi$ and $\pm 4\pi$, corresponding to the local OAM of $\pm 1$ and $\pm 2$, respectively. $|p_+\rangle$ and $|d_+\rangle$ span the pseudospin-up subspace, whereas $|p_-\rangle$ and $|d_-\rangle$ span the pseudospin-down subspace. Although there are six sites in a unit cell, the $\boldsymbol{k}\cdot\boldsymbol{p}$ Hamiltonian based on the four states $(|p_+\rangle \quad |d_+\rangle \quad |p_-\rangle \quad |d_-\rangle) \equiv (|1\rangle \quad |2\rangle \quad |3\rangle \quad |4\rangle)$ can capture the topological property of the system [18, 31]:

$$H(\vec{k}) = \begin{bmatrix} M + Bk^2 & Ak_+ & Ck_-^2 & 0 \\ A^*k_- & -M - Bk^2 & 0 & Ck_+^2 \\ Ck_+^2 & 0 & M + Bk^2 & Ak_- \\ 0 & Ck_-^2 & A^*k_+ & -M - Bk^2 \end{bmatrix}, \tag{2}$$

where $B = a^2 t_{\text{out}}/4$, $M = t_{\text{in}} - t_{\text{out}}$ and $C = a^2 t_{\text{out}}/8$ are real, $A = iat_{\text{out}}/2$ is purely imaginary, and $k_\pm = k_x \pm ik_y$which are measured from the Γ point. It can be checked straightforwardly that $H_{12}(\vec{k}) = Ak_+$ and $H_{13}(\vec{k}) = Ck_-^2$ satisfy the $C_{6v}$ symmetry.

For the ring cavity of a circular shape, the polar coordinate is convenient

$$\begin{aligned} k_x &= \cos\phi\, k_r - \sin\phi\, k_\phi \\ k_y &= \sin\phi\, k_r + \cos\phi\, k_\phi \end{aligned} \tag{3}$$

where $\phi$ is the azimuthal angle with respect to the center of the TRC laser. It is readily seen that the momentum operators $(k_x \quad k_y) = (-i\partial_x \quad -i\partial_y)$ and $(k_r \quad k_\phi) = \left[-i\partial_r \quad (1/r)\left(-i\partial_\phi\right)\right]$ satisfy Eq. (3). One has $k_\pm = (-i\partial_x) \pm i\left(-i\partial_y\right) = \exp(\pm i\phi)\left[(-i\partial_r) \pm i(1/r)\left(-i\partial_\phi\right)\right]$.

The off-diagonal $2 \times 2$ matrices in Eq. (2) only contain high-order terms of $k$, which do not influence the topological property. Therefore, we can treat separately the two pseudospin sectors, and consider the coupling between them when necessary. The Hamiltonian for the pseudospin-up sector is rewritten as

$$H_+ = H_0 + \Delta H_+ \tag{4}$$

with

$$\begin{aligned} H_0 &= \begin{bmatrix} M & Ae^{i\phi}k_r \\ A^*e^{-i\phi}k_r & -M \end{bmatrix}, \\ \Delta H_+ &= \begin{bmatrix} B\left(k_r^2 + k_\phi^2 - \frac{1}{r}\partial_r\right) & iAe^{i\phi}k_\phi \\ -iA^*e^{-i\phi}k_\phi & -B\left(k_r^2 + k_\phi^2 - \frac{1}{r}\partial_r\right) \end{bmatrix}, \end{aligned} \tag{5}$$

where $\Delta H_+$ is taken as a perturbation. For a circular TRC laser, it is natural to change the basis as

$$(|p_+\rangle \quad |d_+\rangle) \to (e^{i\phi}|p_+\rangle \quad |d_+\rangle), \tag{6}$$

and under this new basis the Hamiltonian $H_0$ is recast into

$$H_0 = \begin{bmatrix} M & Ak_r \\ A^*k_r & -M \end{bmatrix} \equiv H_{1D}. \tag{7}$$

Now we proceed to analyze the whole system with the topological PhC at the center which is cladded by a trivial PhC. The mass in the one-dimensional Dirac Hamiltonian (7) takes opposite signs in the topological and trivial PhCs: $M(r < R_0) = +M_<$ and $M(r > R_0) = -M_>$ with $M_{<,>} > 0$. This leads to the Jackiw-Rebbi zero-energy soliton solution localized at the interface [32], from which the TRC modes emerge as shown in the following. With the radius dependences of the quantities in Eq. (7), the Hamiltonian is given explicitly by

$$H_{1D}=\begin{bmatrix} M(r) & A(r)k_r \\ A^*(r)k_r & -M(r)\end{bmatrix}=\begin{bmatrix} M(r) & -|A(r)|\partial_r \\ |A(r)|\partial_r & -M(r)\end{bmatrix}. \tag{8}$$

For $r<R_0$, the trial wave function can be written as

$$\psi^{<}(r)=\begin{bmatrix}\psi_1^{<} \\ \psi_2^{<}\end{bmatrix}e^{(r-R_0)\kappa_<} \tag{9}$$

where the inverse decay length $\kappa_<$ should satisfy

$$\begin{vmatrix} M_< - E & -|A_<|\kappa_< \\ |A_<|\kappa_< & -M_< - E\end{vmatrix}=0 \tag{10}$$

with an eigenvalue $E$. This gives

$$\kappa_< = \pm\frac{\sqrt{M_<^2-E^2}}{|A_<|}. \tag{11}$$

The physical solution with $\psi(r\to 0)\to 0$ requires $\kappa_<>0$ and $M_<>E$, and the wave function

$$\psi_1^{<}=\frac{|A_<|\kappa_<}{M_<-E}\psi_2^{<}=\frac{\sqrt{M_<^2-E^2}}{M_<-E}\psi_2^{<} \tag{12}$$

Similarly, for $r>R_0$, the trial wave function can be written as

$$\psi^{>}(r)=\begin{bmatrix}\psi_1^{>} \\ \psi_2^{>}\end{bmatrix}e^{-(r-R_0)\kappa_>} \tag{13}$$

with

$$\begin{vmatrix} -M_> - E & |A_>|\kappa_> \\ -|A_>|\kappa_> & M_> - E\end{vmatrix}=0 \tag{14}$$

and

$$\kappa_> = \pm\frac{\sqrt{M_>^2-E^2}}{|A_>|}. \tag{15}$$

The physical solution with $\psi(r\to\infty)\to 0$ requires $\kappa_>>0$ and $M_>>E$, and

$$\psi_1^{>}=\frac{|A_>|\kappa_>}{M_>+E}\psi_2^{>}=\frac{\sqrt{M_>^2-E^2}}{M_>+E}\psi_2^{>}. \tag{16}$$

The continuation of the wave function at the interface between the topological and trivial PhCs requires

$\psi^{<}(r=R_0)=\psi^{>}(r=R_0)$, such that

$$\frac{\sqrt{M_<^2-E^2}}{M_<-E}=\frac{\sqrt{M_>^2-E^2}}{M_>+E}. \tag{17}$$

This results in $E=0$ and the whole wave function is given by

$$\psi(r) = \sqrt{\frac{M_< M_>}{M_<|A_>| + M_>|A_<|}}\begin{bmatrix}1\\1\end{bmatrix}\begin{cases} e^{\frac{M_<}{|A_<|}(r-R_0)}, & r < R_0 \\ e^{-\frac{M_>}{|A_>|}(r-R_0)}, & r > R_0 \end{cases} \equiv \begin{bmatrix}1\\1\end{bmatrix} F(r) \tag{18}$$

with $\int_0^\infty F(r)^2\, dr = 1/2$, where a term proportional to $\exp(-M_< R_0/|A_<|)$ is considered infinitesimal and thus neglected. The eigenstate is

$$|+\rangle = \left(e^{i\phi}|p_+\rangle + |d_+\rangle\right)F(r) = (|p_+\rangle \quad |d_+\rangle)\begin{pmatrix} e^{i\phi} \\ 1 \end{pmatrix} F(r). \tag{19}$$

Now we consider the perturbation $\Delta H_+$ keeping in mind that the state is localized close to the interface $r = R_0$

$$\begin{aligned} &\langle +|\Delta H_+|+\rangle \\ &= \int_0^\infty dr(e^{-i\phi} \quad 1)F(r)\begin{bmatrix} B(r)\left(k_r^2 + k_\phi^2 - \frac{1}{r}\partial_r\right) & iA(r)e^{i\phi}k_\phi \\ -iA(r)^* e^{-i\phi}k_\phi & -B(r)\left(k_r^2 + k_\phi^2 - \frac{1}{r}\partial_r\right) \end{bmatrix}\begin{pmatrix} e^{i\phi} \\ 1 \end{pmatrix} F(r) \\ &\approx \left(\int_0^\infty |A|(r)F(r)^2 dr\right)\left(k_\phi + e^{-i\phi}k_\phi\left(e^{i\phi}\right)\right) = \frac{M_<|A_>|^2 + M_>|A_<|^2}{R_0(M_<|A_>| + M_>|A_<|)}\left(-i\partial_\phi + \frac{1}{2}\right) \\ &\equiv \hbar\omega_0\left(-i\partial_\phi + \frac{1}{2}\right), \end{aligned} \tag{20}$$

where $k_\phi = (1/R_0)\left(-i\partial_\phi\right)$ is taken into account, and the term $Bk_\phi^2$ is neglected as a high-order correction since $1/\kappa_<, 1/\kappa_> \ll R_0$. With the hopping integrations for the topological and trivial PhCs and the size of TRC $R_0 = 5a$, one obtains the estimate $\hbar\omega_0 \approx 0.39 t_0$.

It is straightforward to see that the perturbed states are given by $e^{im\phi}$ with $m$ being integers with eigenvalues $\hbar\omega = \hbar\omega_0(m + 1/2)$. We point out that the offset of 1/2 in the energy spectrum originates from the Dirac physics in the present system.

Similarly, we can derive the zero-energy Jackiw-Rebbi state in the pseudospin-down sector

$$|-\rangle = \left(e^{-i\phi}|p_-\rangle + |d_-\rangle\right)F(r) = (|p_-\rangle \quad |d_-\rangle)\begin{pmatrix} e^{-i\phi} \\ 1 \end{pmatrix} F(r), \tag{21}$$

and the perturbed states $e^{-im\phi}$, which are the TR counterparts of those in the pseudospin-up sector, with the same eigenvalues.

Keeping in mind that the above discussion is around the center of frequency band gap $\omega_c$, we arrive at

$$\begin{gathered}\omega_+(q) = \omega_c + q\omega_0, \\ |q,+\rangle = e^{iq\phi}\left(e^{\frac{i\phi}{2}}|p_+\rangle + e^{-\frac{i\phi}{2}}|d_+\rangle\right)F(r), \\ \omega_-(q) = \omega_c - q\omega_0, \\ |q,-\rangle = e^{iq\phi}\left(e^{-\frac{i\phi}{2}}|p_-\rangle + e^{\frac{i\phi}{2}}|d_-\rangle\right)F(r),\end{gathered} \tag{22}$$

with $q = \cdots, -5/2, -3/2, -1/2, 1/2, 3/2, 5/2, \cdots$ (see also [33]).

The two sets of eigen wave functions in Eq. (22) are the helical interface states in a circular TRC, where the pseudospin-up (-down) states carry the counterclockwise (clockwise) group velocity $\delta\omega_+(q)/\delta q > 0$ ($\delta\omega_-(q)/\delta q < 0$), as the manifestation of *spin-momentum locking* in topological interface states with TR symmetry [6, 7, 18]. The spectrum (22) is particle-hole symmetric with respect to the center of frequency band gap $\omega_c$ due to the chiral symmetry of Hamiltonian (1), which can also be seen in the numerical results shown below. For simplicity, we use $\omega_q \equiv \omega_+(q) = \omega_-(-q)$ in the following discussions, considering that $|q,+\rangle$ and $|-q,-\rangle$ are degenerate in frequency as guaranteed by TR symmetry.

### B. Numerical analyses and simulations based on COMSOL

We can also analyze the finite system by diagonalizing Hamiltonian (1) numerically: $H\psi = \hbar\omega\psi$ with the wave function $\psi = (a_1 \quad \cdots \quad a_N)^T$. In Fig. 3(a), the eigenvalues for the TRC modes are superimposed on the dispersions of the two PhCs side-by-side in a long ribbon geometry, with the latter giving the relevant scales: $\hbar\omega_{\pm 7/2} = \pm 1.20t_0$, $\hbar\omega_{\pm 5/2} = \pm 0.87t_0$, $\hbar\omega_{\pm 3/2} = \pm 0.53t_0$ and $\hbar\omega_{\pm 1/2} = \pm 0.18t_0$ with the particle-hole symmetry. The frequency separation $\hbar\omega_0 \approx 0.34t_0$ agrees satisfactorily with the estimate derived analytically from the Dirac equation. There is a small frequency splitting in the TRC $q = \pm 3/2$ denoted by the light-colored circles in the dispersion diagram in Fig. 3(a), which will be discussed later. We display in Figs. 3(b) and (c) [Figs. 3(d) and (e)] the real and imaginary parts of the up-pseudospin eigen wave function for $|-1/2,+\rangle$ ($|1/2,+\rangle$) at $\hbar\omega_{q=-1/2} = -0.18t_0$ ($\hbar\omega_{q=+1/2} = 0.18t_0$). As can be read from the phase change along the closed TRC and parity of the wave function with respect to the device center, these two TRC modes carry global OAM $L = +1$ and $L = +2$, respectively. All of these match completely with Eq. (22).

Simulations based on COMSOL also confirm the above results derived from the TB model. In order to perform COMSOL simulation, we take PhCs of dielectric cylinders $a = 649.5$nm, $d = 128.3$nm, $\epsilon = 11.7\epsilon_0$, $R_< = 242.5$nm for the topological PhC and $R_> = 190.5$nm for the trivial PhC, with $R_0 = 5a$ the same as in the TB model. The common frequency band gap lies in the frequency window $f = 226.7$~$254.6$THz.

We display in Figs. 4(a)-(d) the real and imaginary parts of the out-of-plane electric field obtained by COMSOL for the two TRC modes at frequencies $f_{q=-1/2} = 237.3$THz and $f_{q=1/2} = 239.8$THz, which reside just below and above the center of the common bulk frequency band gap of the topological and trivial PhCs. The state given in Figs. 4(a) and (b) [Figs. 4(c) and (d)] is specified by a *global* OAM of $L = +1$ ($L = +2$) with respect to the center of the system as can be read from the field distributions, which corresponds to the eigenstate $|-1/2,+\rangle$ ($|1/2,+\rangle$) obtained by the TB model shown in Figs. 3(b) and (c) [Figs. 3(d) and (e)].

We have also found doubly degenerate eigenstates at $f_{q=-7/2} = 227.5$THz, $f_{q=-5/2} = 230.6$THz, $f_{q=+5/2} = 246.1$THz and $f_{q=+7/2} = 248.0$THz. One notices that the frequencies of TRC modes are not separated very equally, which can be attributed to the next n.n. couplings among dielectric cylinders neglected in the TB model.

### C. Coupling between two pseudospin sectors

From COMSOL simulations, the four states specified by $q = \pm 3/2$ behave in a way slightly different from other TRC modes discussed above. The TR pairs couple to each other and form two standing waves at

$f^{(1)}_{q=-3/2} = 232.7\text{THz}$ and $f^{(2)}_{q=-3/2} = 235.3\text{THz}$, and $f^{(1)}_{q=3/2} = 241.1\text{THz}$ and $f^{(2)}_{q=3/2} = 244.9\text{THz}$. In order to understand this feature, we consider the off-diagonal $2\times 2$ matrices $H_{\text{CP}} = \text{diag}\{Ck_-^2, Ck_+^2\}$ in Eq. (2), which couple the two pseudospin sectors. Taking this coupling as a perturbation to the helical ring-cavity modes, the perturbed energy of a TR pair $\langle q,+|H_{\text{CP}}|-q,-\rangle$ is given by

$$\begin{aligned}
&\int_0^\infty rdr\int_0^{2\pi} d\phi\left(e^{-i\left(q+\frac{1}{2}\right)\phi} \quad e^{-i\left(q-\frac{1}{2}\right)\phi}\right)F(r)\begin{bmatrix} Ck_-^2 & 0 \\ 0 & Ck_+^2\end{bmatrix}\begin{pmatrix} e^{-i\left(q+\frac{1}{2}\right)\phi} \\ e^{-i\left(q-\frac{1}{2}\right)\phi}\end{pmatrix}F(r) \\
&\qquad = \int_0^\infty rdr\int_0^{2\pi} d\phi\left[\left(e^{-i\left(q+\frac{1}{2}\right)\phi}F(r)Ck_-^2 e^{-i\left(q+\frac{1}{2}\right)\phi}F(r)\right)\right. \\
&\qquad\quad \left.+\left(e^{-i\left(q-\frac{1}{2}\right)\phi}F(r)Ck_+^2 e^{-i\left(q-\frac{1}{2}\right)\phi}F(r)\right)\right]
\end{aligned} \tag{23}$$

where $k_\pm^2 = e^{\pm 2i\phi}\left(-\partial_r^2 + \frac{1}{r^2}\partial_\phi^2 + \frac{1}{r}\partial_r \mp \frac{2i}{r}\partial_\phi\partial_r \pm \frac{2i}{r^2}\partial_\phi\right)$. It is clear that the above integration is non-zero for $q = \pm 3/2$, which mixes the pair of pseudospin-up and -down states. For other values of $q$, including $q = \pm 1/2$, the two pseudospin sectors remain intact.

## III. LASING BEHAVIORS of TRC MODES

### A. Complex eigen spectrum

Now we consider the dynamics $i\hbar\dot{\psi} = H\psi$ with $H$ given in Eq. (1) when gains and loss are added. The full equation of motion is given by [26, 31, 34, 35]

$$i\hbar\dot{a}_n = \sum\nolimits_{\langle n.n.\rangle} t_{nm}a_m + \left(\frac{igP_n}{1+|a_n|^2/I_{\text{sat}}} - i\gamma\right)a_n, \tag{24}$$

where $gP_n$ is the gain arranged along the interface [$P_n = 1$ if site $n$ is on the interface, and 0 otherwise; see Fig. 1(b)], $I_{\text{sat}}$ determines the saturation value of the wave function, and $\gamma > 0$ represents possible losses in the system, including spontaneous leaks due to that the topological interface states lie above the light line.

In order to check the time evolution of the system including the non-Hermitian part of Hamiltonian, we start from an initial state $a_{kn}(t=0)$ proportional to the $k$th eigen wave function $\{\tilde{a}_{kn}\}$ of the Hermitian

Hamiltonian,

$$i\hbar\dot{a}_{kn}(t)=\sum\nolimits_{\langle n.n.\rangle}t_{nm}a_{km}(t)+\left(\frac{igP_n}{1+|a_{kn}|^2/I_{\text{sat}}}-i\gamma\right)a_{kn}(t). \tag{25}$$

With $a_{kn}(t)=\tilde{a}_{kn}W_k(t)$ where $W_k(0)=1$, one has

$$\begin{aligned} &i\hbar\dot{W}_k(t)\sum\nolimits_n\tilde{a}_{kn}^*\,\tilde{a}_{kn} \\ &=W_k(t)\sum\nolimits_{n,m}\tilde{a}_{kn}^*t_{nm}\tilde{a}_{km}+W_k(t)\sum\nolimits_n\tilde{a}_{kn}^*\left(\frac{igP_n}{1+|\tilde{a}_{kn}|^2|W_k(t)|^2/I_{\text{sat}}}-i\gamma\right)\tilde{a}_{kn}. \end{aligned} \tag{26}$$

Taking $\sum_n\tilde{a}_{kn}^*\tilde{a}_{kn}=1$ and $\hbar\omega_{k\text{R}}=\sum_{n,m}\tilde{a}_{kn}^*t_{nm}\tilde{a}_{km}$, we obtain

$$\dot{W}_k(t)/W_k(t)=-i\omega_{k\text{R}}+\frac{1}{\hbar}\sum\nolimits_n|\tilde{a}_{kn}|^2\left(\frac{gP_n}{1+|\tilde{a}_{kn}|^2|W_k(t)|^2/I_{\text{sat}}}-\gamma\right). \tag{27}$$

Therefore, the imaginary part of the eigenvalue

$$\omega_{k\text{I}}(t)=\frac{1}{\hbar}\sum\nolimits_n|\tilde{a}_{kn}|^2\left(\frac{gP_n}{1+|\tilde{a}_{kn}|^2|W_k(t)|^2/I_{\text{sat}}}-\gamma\right) \tag{28}$$

gives the growing rate of the $k$th eigenstate.

With gains assigned at the interface between the two PhCs [see Fig. 1(b)] such that $P_n=1$ for site $n$ on the interface and 0 otherwise, we obtain the following estimate on the imaginary part of the eigenvalue

$$\begin{aligned} \omega_{k\text{I}}(t)&=-\frac{\gamma}{\hbar}+\frac{g}{\hbar}\sum\nolimits_n'\frac{|\tilde{a}_{kn}|^2}{1+|\tilde{a}_{kn}|^2|W_k(t)|^2/I_{\text{sat}}} \\ &=-\frac{\gamma}{\hbar}+\frac{gI_{\text{sat}}}{\hbar|W_k(t)|^2}\left(N'-\frac{|W_k(t)|^2}{I_{\text{sat}}}\sum\nolimits_n'\frac{1}{|\tilde{a}_{kn}|^2+I_{\text{sat}}/|W_k(t)|^2}\right) \\ &\leq-\frac{\gamma}{\hbar}+\frac{gI_{\text{sat}}N'}{\hbar|W_k(t)|^2}\left[1-\frac{N'}{N'+(|W_k(t)|^2/I_{\text{sat}})\sum_n'|\tilde{a}_{kn}|^2}\right] \end{aligned} \tag{29}$$

where $\Sigma_n'$ is the summation taken over interface sites where gains are assigned and $N'=\sum_n'1$, the inequality of arithmetic and harmonic means $(\sum_n 1/c_n)/N\geq N/(\sum_n c_n)$, with $c_n>0,\forall n>0$, is considered. Therefore, among eigenstates with the same norm, a state with the most uniform amplitude along the interface takes the largest imaginary eigenvalue, and grows most quickly with time. Moreover, as can be read from Eq. (29), the maximal value itself increases with $\sum_n'|\tilde{a}_{kn}|^2$, indicating that the more concentrated the wave function at the interface the larger the imaginary eigenvalue.

### B. Uniformity of TRC mode

It is clear that the four TRC modes $|\pm 1/2, \pm\rangle$ residing closest to the center of the frequency band gap are most concentrated at the interface, since the decay length is proportional to the inversed frequency difference measured from the band edge. In the following, we discuss how the cavity shape influences the uniformity of the TRC modes along the azimuthal direction. Because the shape effect is TR symmetric, it certainly does not induce coupling between the two pseudospin channels, which allows us to start from the Hamiltonian

$$H_{⬢}(\vec{k}) = \begin{bmatrix} M + Bk_{⬢}^2 & Ak_{⬢+} & 0 & 0 \\ A^*k_{⬢-} & -M - Bk_{⬢}^2 & 0 & 0 \\ 0 & 0 & M + Bk_{⬢}^2 & Ak_{⬢-} \\ 0 & 0 & A^*k_{⬢+} & -M - Bk_{⬢}^2 \end{bmatrix}. \tag{30}$$

where the parameters $M, B$ and $A$ are determined by the wave functions inside the unit cell as given in Eq. (2), which form the basis and do not change with the cavity shape, and the symbol ⬢ refers to the hexagonal shape of ring cavity

$$\begin{gathered} R(\phi) = R_0/f(\phi) \\ f(\phi) = \cos\left(\phi - \frac{\pi}{6} - \frac{\pi}{3}n\right), \phi \in \left[\frac{\pi}{3}n, \frac{\pi}{3}(n+1)\right), n = 0,1,2,3,4,5 \end{gathered} \tag{31}$$

with $\phi = 0$ starting from the right corner of the hexagonal cavity as shown in Fig. 1.

We first focus on the pseudospin-up sector, for which Hamiltonian (30) is rewritten as

$$H_{+⬢} = H_{0⬢} + \Delta H_{+⬢} \tag{32}$$

with

$$\begin{gathered} H_{0⬢} = \begin{bmatrix} M & Ae^{i\phi}k_r \\ A^*e^{-i\phi}k_r & -M \end{bmatrix}, \\ \Delta H_{+⬢} = \begin{bmatrix} B\left(k_r^2 + k_{\phi⬢}^2 - \frac{1}{r_{⬢}}\partial_r\right) & iAe^{i\phi}k_{\phi⬢} \\ -iA^*e^{-i\phi}k_{\phi⬢} & -B\left(k_r^2 + k_{\phi⬢}^2 - \frac{1}{r_{⬢}}\partial_r\right) \end{bmatrix} \end{gathered} \tag{33}$$

where $H_{0⬢}$ gives birth to the Jackiw-Rebbi zero-energy soliton due to the opposite signs of mass in the photonic Dirac dispersions inside and outside the closed interface, as revealed above for the circular cavity, and $\Delta H_+$ is taken as a perturbation. In the following calculations, small differences in quantities $A$ and $B$ between the two PhCs are omitted for clarity. While the operator $k_r = -i\partial_r$ remains unchanged, $k_{\phi\text{⬢}}$ changes with the cavity shape since it is associated with the Jackiw-Rebbi interface state $k_{\phi\text{⬢}} = [f(\phi)/R_0](-i\partial_\phi)$. With the similar procedure for the circular cavity, we obtain the eigenstates of $H_{0\text{⬢}}$

$$
\begin{aligned}
|+\rangle_{\text{⬢}} &= (|p_+\rangle \quad |d_+\rangle)\begin{pmatrix} e^{i\phi} \\ 1 \end{pmatrix} F(r,\phi) \\
&= (|p_+\rangle \quad |d_+\rangle)\begin{pmatrix} e^{i\phi} \\ 1 \end{pmatrix} \sqrt{\frac{M_< M_>}{|A|(M_< + M_>)}} \begin{cases} e^{\frac{M_<}{|A|}(r-R(\phi))}, & r < R(\phi) \\ e^{-\frac{M_>}{|A|}(r-R(\phi))}, & r > R(\phi) \end{cases}
\end{aligned} \tag{34}
$$

with $\int_0^\infty F(r,\phi)^2\, dr = 1/2$.

Next, we consider the perturbation from $\Delta H_{+\text{⬢}}$ to the wave function $|+\rangle_{\text{⬢}}$, noticing $k_{\phi\text{⬢}} = [1/R(\phi)](-i\partial_\phi) = [f(\phi)/R_0](-i\partial_\phi)$ since the Jackiw-Rebbi soliton is localized at the interface $r = R(\phi)$:

$$
\begin{aligned}
&{}_{\text{⬢}}\langle +|\Delta H_{+\text{⬢}}|+\rangle_{\text{⬢}} \\
&= \int_0^\infty dr (e^{-i\phi} \quad 1) F(r,\phi) \begin{bmatrix} B\left(k_r^2 + k_{\phi\text{⬢}}^2 - \frac{1}{r_{\text{⬢}}}\partial_r\right) & iAe^{i\phi} k_{\phi\text{⬢}} \\ -iA^* e^{-i\phi} k_{\phi\text{⬢}} & -B\left(k_r^2 + k_{\phi\text{⬢}}^2 - \frac{1}{r_{\text{⬢}}}\partial_r\right) \end{bmatrix} \begin{pmatrix} e^{i\phi} \\ 1 \end{pmatrix} F(r,\phi) \\
&\approx \int_0^\infty dr (e^{-i\phi} \quad 1) F(r,\phi) \begin{bmatrix} B\left(k_r^2 - \frac{1}{R(\phi)}\partial_r\right) & iAe^{i\phi}\frac{1}{R(\phi)}(-i\partial_\phi) \\ -iA^* e^{-i\phi}\frac{1}{R(\phi)}(-i\partial_\phi) & -B\left(k_r^2 - \frac{1}{R(\phi)}\partial_r\right) \end{bmatrix} \begin{pmatrix} e^{i\phi} \\ 1 \end{pmatrix} F(r,\phi) \\
&= \left(\int_0^\infty F(r,\phi)^2 dr\right)\frac{1}{R_0} f(\phi)\left(\frac{1}{2} - i\partial_\phi\right) + \frac{1}{R(\phi)}\int_0^\infty \frac{-i\partial F(r,\phi)^2}{\partial\phi} dr \\
&= \frac{|A|}{R_0}\left(\frac{1}{2} - i\partial_\phi\right) + (f(\phi) - 1)\frac{|A|}{R_0}\left(\frac{1}{2} - i\partial_\phi\right) \\
&\equiv \Delta H_1 + \Delta H_2.
\end{aligned} \tag{35}
$$

We have already shown that the eigen wave functions of $\Delta H_1 = (|A|/R_0)(1/2 - i\partial_\phi)$ are $\varphi(\phi) = \exp[i(q - 1/2)\phi]$, whereas $\Delta H_2$ can be taken as a perturbation since $|f(\phi) - 1|_{\max} = 1 - \sqrt{3}/2 < 1$. Therefore, performing the first-order perturbation treatment, one has

$$
\begin{aligned}
&\varphi_{\text{⬢}}(\phi) \\
&= e^{i\left(q-\frac{1}{2}\right)\phi} + \sum_{q'\neq q} \frac{\int_0^{2\pi} d\phi' \left[ e^{-i\left(q'-\frac{1}{2}\right)\phi'} [f(\phi')-1] \frac{|A|}{R_0} \left(\frac{1}{2} - i\partial_{\phi'}\right) e^{i\left(q-\frac{1}{2}\right)\phi'} \right]}{\omega_q - \omega_{q'}} e^{i\left(q'-\frac{1}{2}\right)\phi} \\
&= e^{i\left(q-\frac{1}{2}\right)\phi} + q \sum_{q'\neq q} \frac{\int_0^{2\pi} d\phi' [f(\phi')-1] e^{i(q-q')\phi'}}{q-q'} e^{i\left(q'-\frac{1}{2}\right)\phi}.
\end{aligned}
\tag{36}
$$

Taking Eq. (31) into account, one has

$$
\begin{aligned}
&\int_0^{2\pi} d\phi' (f(\phi')-1) e^{i(q-q')\phi'} = \sum_{j=0}^{5} e^{\frac{i(q-q')j\pi}{3}} \int_0^{\frac{\pi}{3}} d\phi' (f(\phi')-1) e^{i(q-q')\phi'} \\
&= \begin{cases} 6\int_0^{\frac{\pi}{3}} d\phi' (f(\phi')-1) e^{i(q-q')\phi'} \neq 0, & \text{for } q-q' = 6n \\ 0, & \text{otherwise} \end{cases}
\end{aligned}
\tag{37}
$$

with $n$ being integer. Finally, we obtain the wave functions:

$$
\begin{aligned}
|q,+\rangle_{\text{⬢}} &= \varphi_{\text{⬢}}(\phi)\left(e^{i\phi}|p_+\rangle + |d_+\rangle\right)F(r,\phi) \\
|q,-\rangle_{\text{⬢}} &= \varphi^*_{\text{⬢}}(\phi)\left(e^{-i\phi}|p_-\rangle + |d_-\rangle\right)F(r,\phi) \\
\varphi_{\text{⬢}}(\phi) &= e^{i\left(q-\frac{1}{2}\right)\phi}\left[1 - q\sum_{n\in Z, n\neq 0} \frac{e^{-i6n\phi}}{n(6n+1)(6n-1)}\right].
\end{aligned}
\tag{38}
$$

The variation in the amplitudes of ring-cavity wave functions is thus proportional to $|q|^2$ for the hexagonal cavity considered in the present work.

The above analyses agree with the numerical results as shown in Figs. 5(a) and (b), where it is clear that the four TRC modes $|\pm 1/2, \pm\rangle$ are most concentrated across the interface, and meanwhile most uniform along the azimuthal direction. As a result, these four TRC modes take the largest imaginary eigenvalue as shown in Fig. 5(c), consistent with Eq. (29).

As can be seen in the full dynamics simulations by numerically integrating Eq. (24) based on the split-step method [31, 36, 37], the four TRC modes $|\pm 1/2, \pm\rangle$ start to take positive imaginary eigen value at the threshold gain $g_{\mathrm{th}} = 2.6\gamma$ and the amplitudes of the wave function increase with time as shown in Fig. 6(a), and grow faster than other modes at larger gain values [see Fig. 6(b) for $g = 8\gamma$]. This feature has a clear topological origin and makes the four modes start lasing before any other modes when gains are introduced along the interface, which is the main result of the present work.

### C. Mode selection by global OAM

It is certainly not practical to use an eigen wave function as the initial state for lasing. Here we show that, taking advantage of the strong competitiveness and conservation of global OAM, one can stimulate selectively the four TRC modes $|\pm 1/2, \pm\rangle$ by assigning initial states which are relatively easy to prepare in experiments. As the first example, we conceive an initial state where small amplitudes are set on the central unit cell of the system and a $2\pi$ phase winding counterclockwise over the six sites, and check the time evolution. As displayed in Fig. 7(a), while intensities on sites in the bulk decrease quickly to zero, intensities on sites close to the interface grow significantly with time. In the saturated state the EM energy flow estimated by the Poynting vector averaged over unit cells transports along the ring cavity counterclockwise, which is exclusively $|-1/2, +\rangle$ as characterized by the pure spectrum in the inset of Fig. 7(a), even though the spectrum weights of TRC modes are vanishingly small in the initial state. This indicates clearly that the state $|-1/2, +\rangle$ possesses a great merit in lasing dynamics compared with other ring-cavity modes with larger absolute value of $q$, consistent with the gain-value dependence of the imaginary part of eigenvalue in Fig. 5. This observation exhibits a completely new *bulk-edge carryover* property which is supported by non-Hermitian topology.

We also try another initial state where finite amplitudes are assigned at the sites which enclose seven unit cells at the center of the TRC laser with $2\pi$ phase increment counterclockwise as shown in Figs. 8(a) and (c). As displayed in Fig. 8(e), the system is driven into the same final state $|-1/2, +\rangle$, in a similar way as shown in Fig. 7.

One can also set finite wave functions with a small amplitude at the six sites at the corners of the hexagonal ring cavity and a $\pi/3$ phase increment counterclockwise as shown in Fig. 8(b), whereby wave functions on all other sites in the system are zero. As can be seen from Fig. 8(d), this initial state is decomposed into the ring-cavity modes $|-1/2,+\rangle$ and $|11/2,+\rangle$ with up pseudospin and $|5/2,-\rangle$ and $|-7/2,-\rangle$ with down pseudospin, and some other bulk states with small weights. As displayed in Fig. 8(f), the amplitudes of the wave functions close to the interface increase sharply due to the energy gains, and large energy exchanges take place among the interface sites at the intermediate stage of time evolution. Finally, the system evolves exclusively into the state $|-1/2,+\rangle$ in a similar way as shown in Fig. 7.

We have also confirmed that the other three TRC modes $|1/2,-\rangle$, $|1/2,+\rangle$ and $|-1/2,-\rangle$ can be achieved by means of the above initial state with phase winding of $-2\pi$, $4\pi$ and $-4\pi$, respectively, where the global OAM clearly plays the crucial role in picking up the destined states. Different from the case shown in Fig. 8, the phase increment between the six corner sites is assigned as $-\pi/3$ as shown in Fig. 9(a), which generates the lasing cavity mode $|1/2,-\rangle$ exclusively. Similarly, the phase increments $2\pi/3$ and $-2\pi/3$ in the initial state lead to the single-mode lasing of $|1/2,+\rangle$ and $|-1/2,-\rangle$, respectively as shown in Figs. 9(b) and 9(c). These four TRC modes $|\pm 1/2,\pm\rangle$ reside at the two sides of the band gap center and share the same amplitude distribution as shown in Fig. 5, thus benefitting optimally from the gain profile.

As demonstrated in very recent experiments, incident circularly-polarized lights can stimulate unidirectional propagation of topological interface modes in setups similar to the present model system [38, 39]. Therefore, it is experimentally feasible to stimulate TRC modes with the specific pseudospin state and the amount of OAM in the present laser device by adjusting judiciously the incident light.

Even using light pumping without phase circulation, the traveling pseudospin-up or pseudospin-down cavity modes should be stimulated to lase, whereas the standing waves formed by two degenerate TR pairs are always suppressed, except for $|\pm 3/2,\pm\rangle$ modes with lifted frequency degeneracy. As can be checked directly in Eq. (22), the traveling states are much more uniform than standing waves along the azimuthal direction, which can avoid possible spatial hole burning.

### D. Robustness of the TRC laser

Varying the gain value, we can map out the behavior of wave function intensity as shown in Fig. 10, which is measured by the squared amplitude of the TRC modes normalized by the number of gain sites. The wave function intensity remains zero for small gains below $g_{\mathrm{th}} = 2.5\gamma$, and then increases linearly with the gain value, which is taken as the lasing event in the present approach where the intensity is considered proportional to the output power. Marvelously, the single-mode lasing is stable up to $g = 370\gamma$, a huge gain value much beyond the scale of Fig. 10, which becomes possible in the TRC laser due to the topological protection as revealed above.

As a benchmark, we have investigated a WGM laser by [34, 35, 40-43],

$$i\hbar\dot{a}_n = \sum\nolimits_{\langle n.n.\rangle} t_{nm}a_m + \left(\frac{ig}{1+|a_n|^2/I_{\mathrm{sat}}} - i\gamma\right)a_n + \Delta_n a_n, \tag{39}$$

as shown schematically in Fig. 11(a) with 54 sites, the same as the number of sites with energy gain in the TRC laser [see Fig. 1(b)], whereas $\Delta_n$ is introduced to capture possible effects of disorder and/or imperfections of PhC. In order to emulate the topological interface state in the TRC laser, we take the site separation in the whispering-gallery cavity as half of the lattice constant of the PhC, such that the hopping integral is $t_{nm} = -2t_0$ in Eq. (39). The dispersion relation is a cosine function as shown in Fig. 11(c) without gains. Adding the effective gain, one obtains complex eigenvalues for the WGM as shown in Fig. 11(d). In a sharp contrast to the TRC modes, all WGMs acquire the same imaginary part in the eigenvalues, which is linearly proportional to the gain value. It is clear that the difference between the TRC laser and the WGM laser stems from the spatial distribution of wave function of the TRC modes.

As seen in Fig. 10, under the similar conditions with the TRC laser discussed above, the WGM laser loses stability at $g = 12\gamma$ ($g_{\mathrm{th}} = 1.0\gamma$) with the initial state shown in Fig. 11(b). The reason for the *intrinsic* instability of WGM is of twofold: first, the imaginary eigenvalue is equal for all WGMs, and secondly, the parabolic behavior makes separations between eigen frequencies very small at the top and bottom of the real cosine dispersion, in sharp contrast to the TRC modes.

Finally, we compare the robustness of TRC laser and WGM laser. A random potential $\Delta_n$ is introduced additionally to Eq. (24) [see Eq. (39)]. For an assigned standard deviation $\sigma$ we test 100 different groups of $\Delta_n$. We can see that the WGM laser is disturbed easily, namely the power spectrum is scattered over many eigenstates as seen in Fig. 12 by the random potential. The robustness of the single-mode lasing in the TRC laser is very clear.

## IV. DISCUSSION

One can also construct the TRC laser with the trivial PhC at the center surrounded by the topological PhC, where the physics addressed above remains unchanged except that the group velocity of the pseudospin-up (-down) channel is clockwise (counterclockwise). The TRC laser can also be realized in PhCs made of dielectric slabs with regular air-hole arrays [44, 45], where transverse electric (TE) modes lase with the ring-cavity modes similar to the ones investigated above. As shown in Fig. 13, the wave functions of the topological interface states for the PhCs made of honeycomb-type arrays of triangular air-holes obtained by COMSOL can be described by the ring-cavity modes given in Eq. (22).

As uncovered in the present work, the four topological interface modes $|\pm 1/2, \pm\rangle$ take larger imaginary parts than other eigen modes, including remaining topological interface modes and other nontopological bulk modes, since they are mostly concentrated at the interface where gain is added, and most uniform along the azimuthal direction. This feature has clear topological origin as revealed from analytical and numerical aspects, which is the main result of the present work. On one hand, our present results set a reference for many possible designs of topological insulator lasers including various shapes of ring cavity, and output ports as well including waveguides attached directly or coupled evanescently to the ring cavity [25-27, 29]. On the other hand, they apply directly to out-of-plane emission [28], where no in-plane buried output port is required. In the out-of-plane emission, the global OAM of the topological interface modes is traced to vorticity, a topological charge of laser vortex, and the pseudospin determines the circular polarization of the laser beam.

Owing to the simple architecture used in the present scheme one can fabricate compact TRC lasers. With the state-of-the-art semiconductor nanotechnologies, one can downscale the lattice constant of the PhC to $\sim$300nm with diameters of cylinders of ~100 nm [46], yielding the total TRC laser of 25μm in linear size, which makes the topological lasing available in visible light. The present PhC platform can be activated by not only light pumping but also electric current injection [29], which is convenient for practical applications.

## V. SUMMARY

We propose a topological ring-cavity laser formed by two photonic crystals distinct in topology obtained by tuning separations between dielectric cylinders in the honeycomb lattice in the designed way, whereby the topological interfacial states constitute the ring-cavity modes and lase when gains are introduced. We find stable single-mode lasing over a wide range of gain value, with the maximal intensity larger than a whispering-gallery-mode counterpart by orders of magnitude. We revealed analytically that the linear dispersion of the topological interfacial states and their topological features are crucial for the by-far stable single-mode lasing. Shining an incident light with circular polarization, one can selectively stimulate lasing modes carrying the desired orbital angular momenta, which can be exploited as a knob to control light propagation and information encoding. The topological ring-cavity laser is commensurate with other Si-based two-dimensional photonic functions, which as a whole can be explored for innovative optic devices.

## ACKNOWLEDGMENTS

This work was supported by the CREST, JST (Core Research for Evolutionary Science and Technology, Japan Science and Technology Agency) (Grant No. JPMJCR18T4), and partially by the Grants-in-Aid for Scientific Research No.17H02913, JSPS (Japan Society of Promotion of Science). We also thank Dr. Ze-Guo Chen for helpful discussion.

**References:**

[1] D. J. Thouless, M. Kohmoto, M. P. Nightingale and M. den Nijs, Phys. Rev. Lett. **49**, 405–408 (1982).

[2] F. D. M. Haldane, Phys. Rev. Lett. **61**, 2015-2018 (1988).

[3] C. L. Kane and E. J. Mele, Phys. Rev. Lett. **95**, 226801 (2005).

[4] B. A. Bernevig, T. L. Hughes and S.-C. Zhang, Science **314**, 1757–1761 (2006).

[5] M. König, S. Wiedmann, C. Brüne, A. Roth, H. Buhmann, L. W. Molenkamp, X.-L. Qi and S.-C. Zhang, Science **318**, 766–770 (2007).

[6] M. Z. Hasan and C. L. Kane, Rev. Mod. Phys. **82**, 3045–3067 (2010).

[7] X.-L. Qi and S.-C. Zhang, Rev. Mod. Phys. **83**, 1057–1110 (2011).

[8] D. Xiao, M.-C. Chang and Q. Niu, Rev. Mod. Phys. **82**, 1959-2007 (2010).

[9] H. Weng, R. Yu, X. Hu, X. Dai and Z. Fang, Adv. Phys. **64**, 227-282 (2015).

[10] F. D. M. Haldane and S. Raghu, Phys. Rev. Lett. **100**, 013904 (2008).

[11] S. Raghu and F. D. M. Haldane, Phys. Rev. A **78**, 033834 (2008).

[12] Z. Wang, Y. D. Chong, J. D. Joannopoulos and M. Soljacic, Phys. Rev. Lett. **100**, 013905 (2008).

[13] Z. Wang, Y. Chong, J. D. Joannopoulos and M. Soljacic, Nature **461**, 772–775 (2009).

[14] M. Hafezi, E. A. Demler, M. D. Lukin and J. M. Taylor, Nat. Phys. **7**, 907–912 (2011).

[15] M. Hafezi, S. Mittal, J. Fan, A. Migdall and J. M. Taylor, Nat. Photon. **7**, 1001–1005 (2013).

[16] M. C. Rechtsman, J. M. Zeuner, Y. Plotnik, Y. Lumer, D. Podolsky, F. Dreisow, S. Nolte, M. Segev and A. Szameit, Nature **496**, 196–200 (2013).

[17] A. B. Khanikaev, S. H. Mousavi, W.-K. Tse, M. Kargarian, A. H. MacDonald and G. Shvets, Nat. Mater. **12**, 233–239 (2013).

[18] L.-H. Wu and X. Hu, Phys. Rev. Lett. **114**, 223901 (2015).

[19] Y. Yang, Y. F. Xu, T. Xu, H.-X. Wang, J.-H. Jiang, X. Hu and Z. H. Hang, Phys. Rev. Lett. **120**, 217401 (2018).

[20] L. Lu, J. D. Joannopoulos and M. Soljacic, Nat. Photon. **8**, 821-829 (2014).

[21] A. B. Khanikaev and G. Shvets, Nat. Photon. **11**, 763-773 (2017).

[22] T. Ozawa, H. M. Price, A. Amo, N. Goldman, M. Hafezi, L. Lu, M. C. Rechtsman, D. Schuster, J. Simon, O. Zilberberg and I. Carusotto, Rev. Mod. Phys. **91**, 015006 (2019).

[23] C. He, X.-C. Sun, X.-P. Liu, M.-H. Lu, Y. Chen, L. Feng and Y.-F. Chen, Proc. Natl. Acad. Sci. USA **113**, 4924–4928 (2016).

[24] Y. Li, Y. Sun, W. Zhu, Z. Guo, J. Jiang, T. Kariyado, H. Chen and X. Hu, Nat.Commun. **9**, 4598 (2018).

[25] B. Bahari, A. Ndao, F. Vallini, A. El Amili, Y. Fainman and B. Kante, Science **358**, 636 (2017).

[26] G. Harari, M. A. Bandres, Y. Lumer, M. C. Rechtsman, Y. D. Chong, M. Khajavikhan, D. N. Christodoulides and M. Segev, Science **359**, eaar4003 (2018).

[27] M. A. Bandres, S. Wittek, G. Harari, M. Parto, J. Ren, M. Segev, D. N. Christodoulides and M. Khajavikhan, Science **359**, eaar4005 (2018).

[28] Z.-K. Shao, H.-Z. Chen, S. Wang, X.-R. Mao, Z.-Q. Yang, S.-L. Wang, X.-X. Wang, X. Hu and R.-M. Ma, Nature Nanotechnology **15**, 67-72 (2020).

[29] Y. Zeng, U. Chattopadhyay, B. Zhu, B. Qiang, J. Li, Y. Jin, L. Li, A. G. Davies, E. H. Linfield, B. Zhang, Y. Chong and Q. J. Wang, Nature **578**, 246-250 (2020).

[30] COMSOL Multiphysics® v. 5.3a. COMSOL AB, Stockholm, Sweden.

[31] L.-H. Wu and X. Hu, Sci. Rep. **6**, 24347 (2016).

[32] R. Jackiw and C. Rebbi, Phys. Rev. D **13**, 3398-3409 (1976).

[33] G. Siroki, P. A. Huidobro and V. Giannini, Phys. Rev. B **96**, 041408 (2017).
[34] J. Butler, D. Ackley and D. Botez, Appl. Phys. Lett. **44**, 293-295 (1984).
[35] S. Wang and H. Winful, Appl. Phys. Lett. **52**, 1774-1776 (1988).
[36] J. A. C. Weideman and B. M. Herbst, SIAM J. Numer. Anal. **23**, 485-507 (1986).
[37] T. R. Taha and M. I. Ablowitz, J. Comput. Phys. **55**, 203-230 (1984).
[38] D. Smirnova, S. Kruk, D. Leykam, E. Melik-Gaykazyan, D.-Y. Choi and Y. Kivshar, Phys. Rev. Lett. **123**, 103901 (2019).
[39] N. Parappurath, F. Alpeggiani, L. Kuipers and E. Verhagen, Sci. Adv. **6**, eaaw4137 (2020).
[40] G. P. Agrawal and N. K. Dutta, *Semiconductor lasers*, (Springer Science & Business Media, Berlin, 2013).
[41] S. McCall, A. Levi, R. Slusher, S. Pearton and R. Logan, Appl. Phys. Lett. **60**, 289-291 (1992).
[42] A. L. Burin, Phys. Rev. E **73**, 066614 (2006).
[43] H. Hodaei, M.-A. Miri, M. Heinrich, D. N. Christodoulides and M. Khajavikhan, Science **346**, 975 (2014).
[44] S. Barik, H. Miyake, W. DeGottardi, E. Waks and M. Hafezi, New J. Phys. **18**, 113013 (2016).
[45] S. Barik, A. Karasahin, C. Flower, T. Cai, H. Miyake, W. DeGottardi, M. Hafezi and E. Waks, Science **359**, 666 (2018).
[46] S. Peng, N. J. Schilder, X. Ni, J. van de Groep, M. L. Brongersma, A. Alù, A. B. Khanikaev, H. A. Atwater and A. Polman, Phys. Rev. Lett. **122**, 117401 (2019).

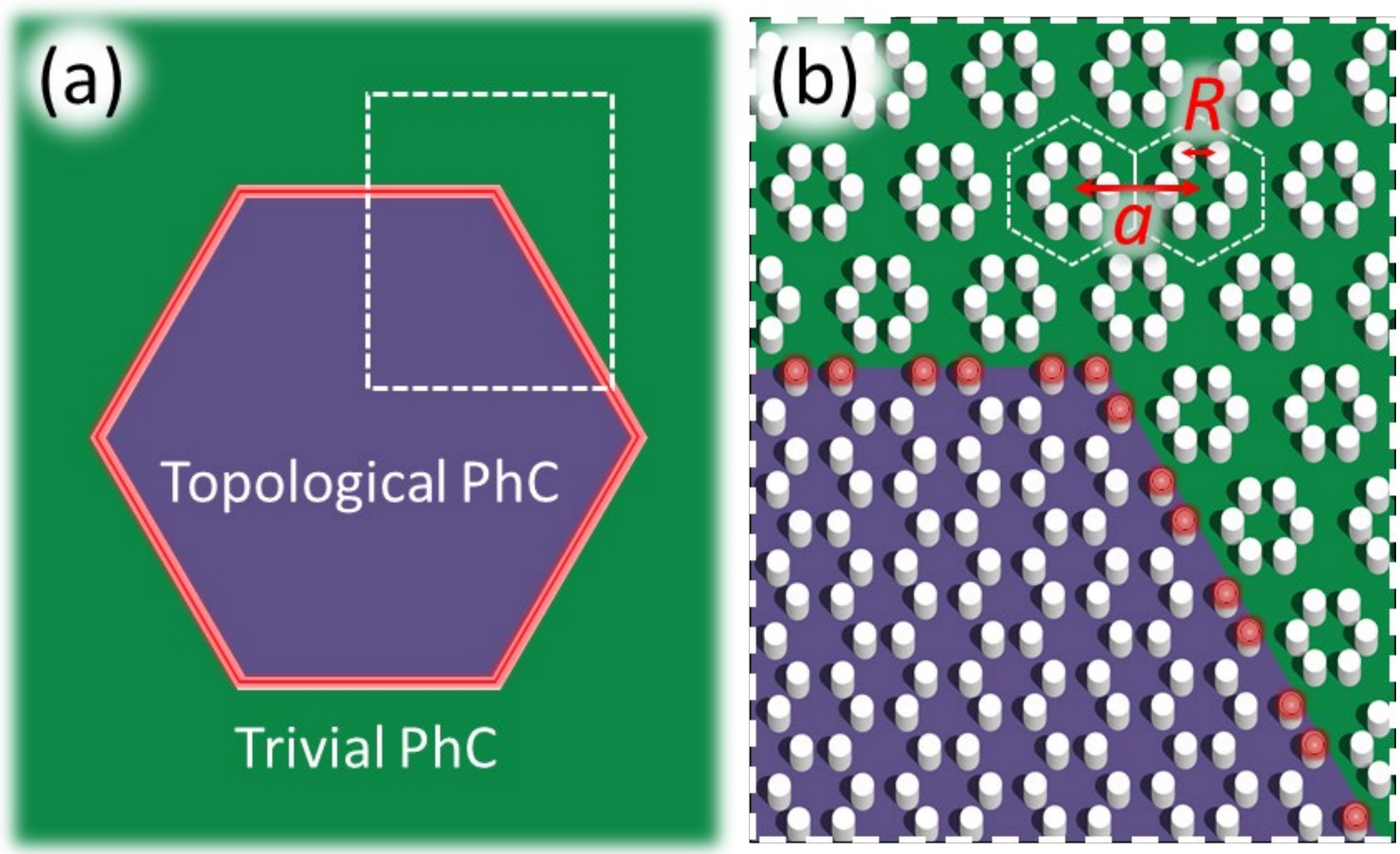

FIG. 1. (a) Schematic of the topological ring-cavity (TRC) laser. (b) Zoom-in picture of the part denoted by dashed line in (a). Gains are introduced on the cylinders in red color on the interface.

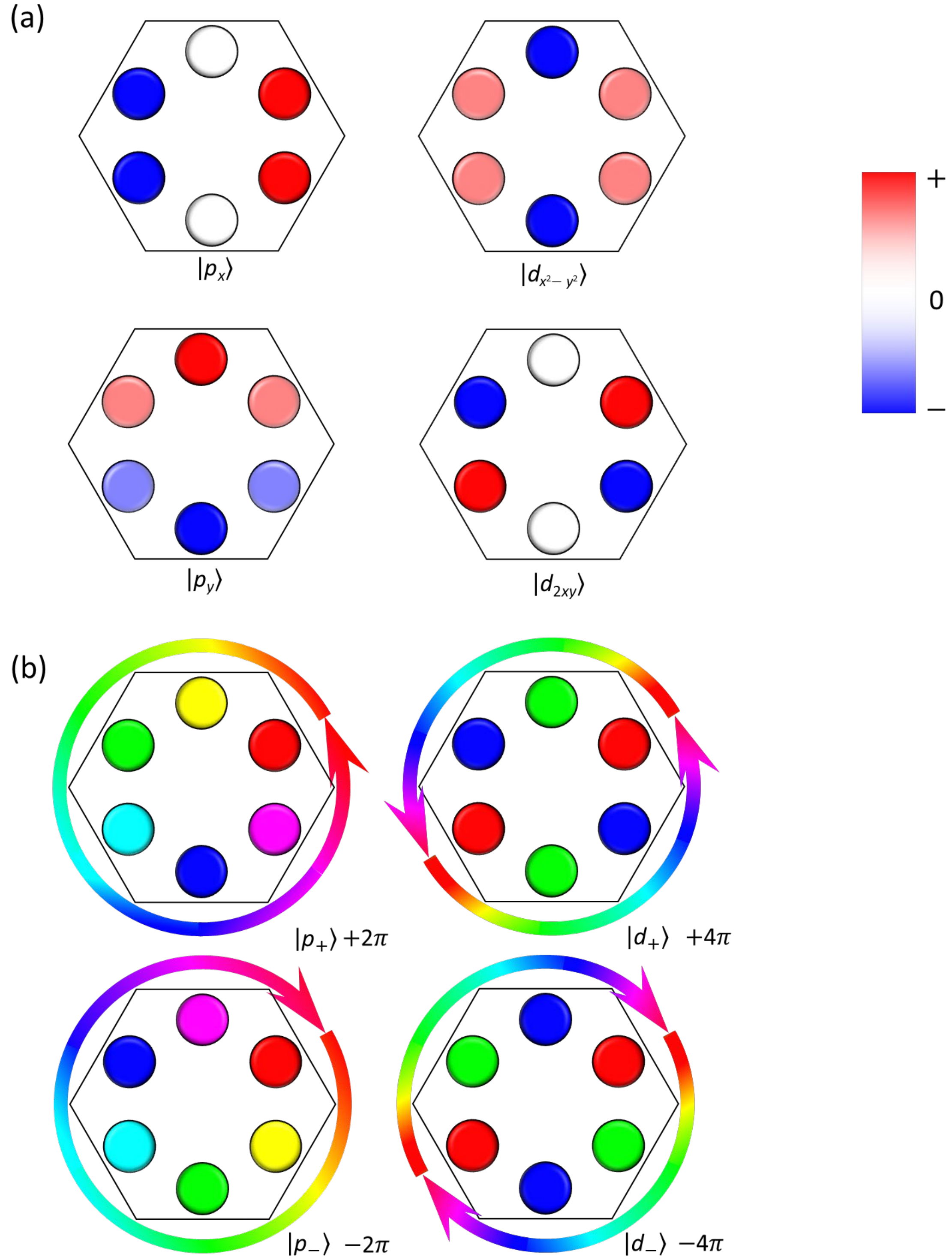

FIG. 2. Eigenstates in the hexagonal unit cell of honeycomb-type lattice. (a) Four eigenstates: $|p_x\rangle/|p_y\rangle$ and $|d_{x^2-y^2}\rangle/|d_{2xy}\rangle$. The color represents the value of z-axis (out-of-plane) electric field with red for positive and blue for negative. (b) Four eigenstates $|p_\pm\rangle = |p_x\rangle \pm i|p_y\rangle$ and $|d_\pm\rangle = |d_{x^2-y^2}\rangle \pm i|d_{2xy}\rangle$ with eigen *local* orbital angular momentum (OAM) $\pm 1$ and $\pm 2$, respectively, where the phase is presented by the colored arrow outside the unit cell. $|p_+\rangle$ and $|d_+\rangle$ form pseudospin-up states, whereas $|p_-\rangle$ and $|d_-\rangle$ form pseudospin-down states.

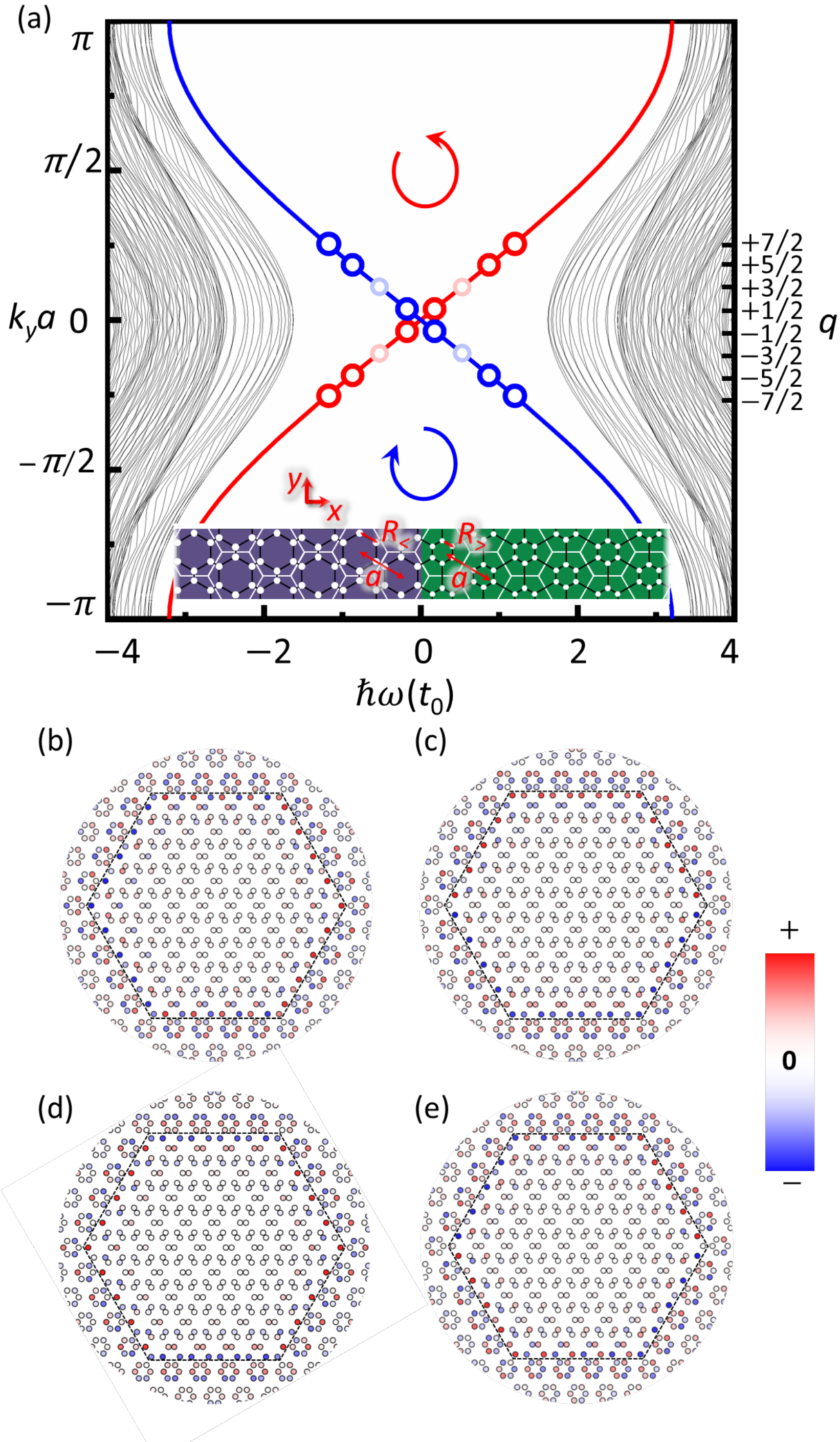

FIG. 3. (a) Dispersion relation of the TRC modes superimposed on the band structure of a ribbon system shown in the inset. (b), (c) Real and imaginary parts of the up-pseudospin TRC mode $|-1/2,+\rangle$ (= Re + $i$Im) of the eigenvalue $\hbar\omega = -0.18t_0$, where $t_0(> 0)$ is the energy unit (see text). (d), (e) Same as (b), (c) except for $|1/2,+\rangle$ (= Re + $i$Im), $\hbar\omega = 0.18t_0$. From the field distribution, one can assign the *global* OAM +1 and +2 to the mode in (b), (c) and (d), (e), respectively.

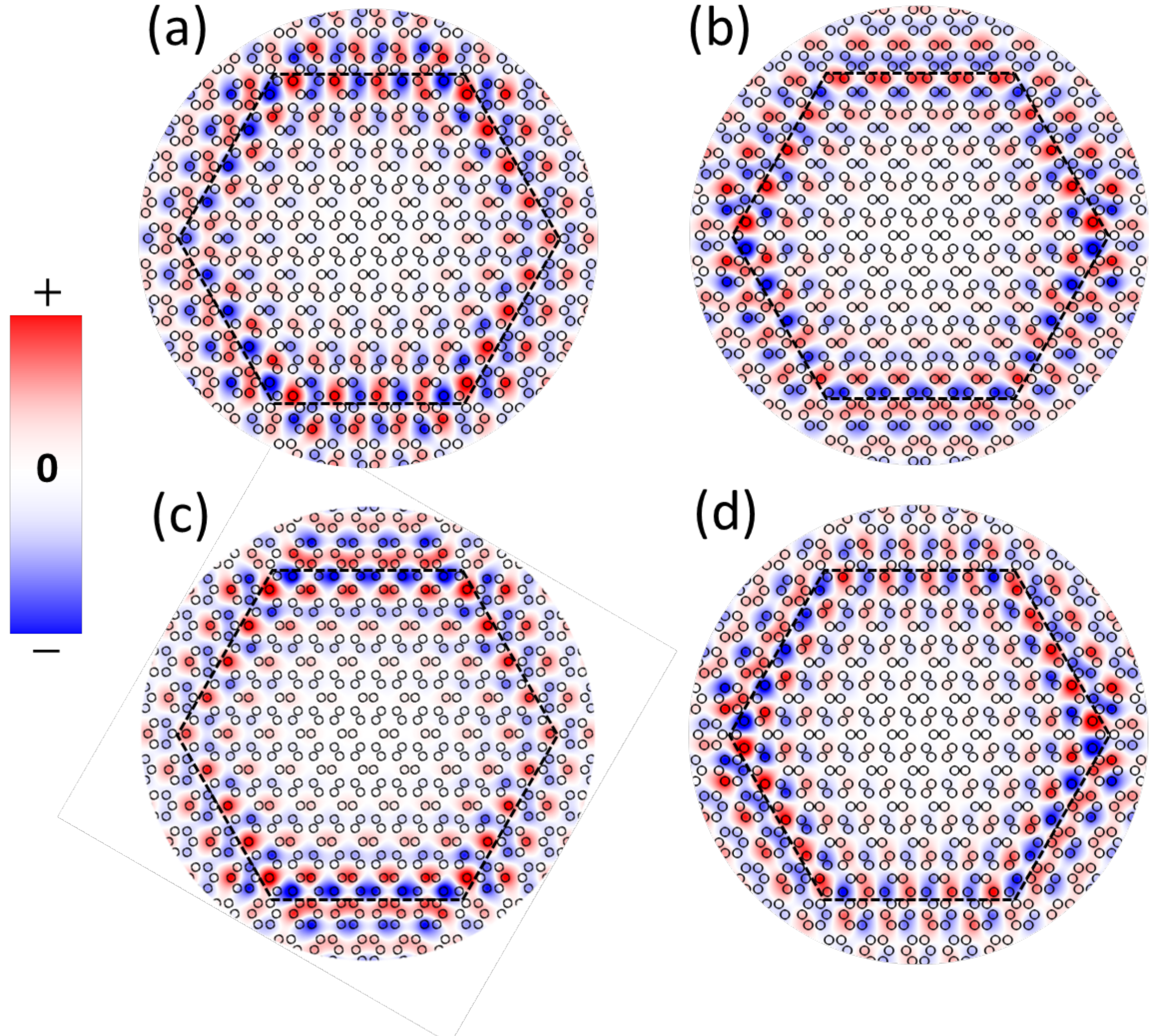

FIG. 4. Cavity eigen wave function obtained by COMSOL, where positions of dielectric cylinders in the two photonic crystals (PhCs) are adjusted to produce the hopping integrals of the TB model. (a), (b) Real and imaginary parts of the up-pseudospin TRC mode $|-1/2,+\rangle$ $(=\mathrm{Re}+i\mathrm{Im})$ of $f=237.3$THz. (c), (d) Same as (a) and (b) except for $|1/2,+\rangle$ $(=\mathrm{Re}+i\mathrm{Im})$, $f=239.8$THz. From the field distribution, one can assign the *global* OAM $+1$ and $+2$ to the mode in (a), (b) and (c), (d), respectively, which match those given in Fig. 3.

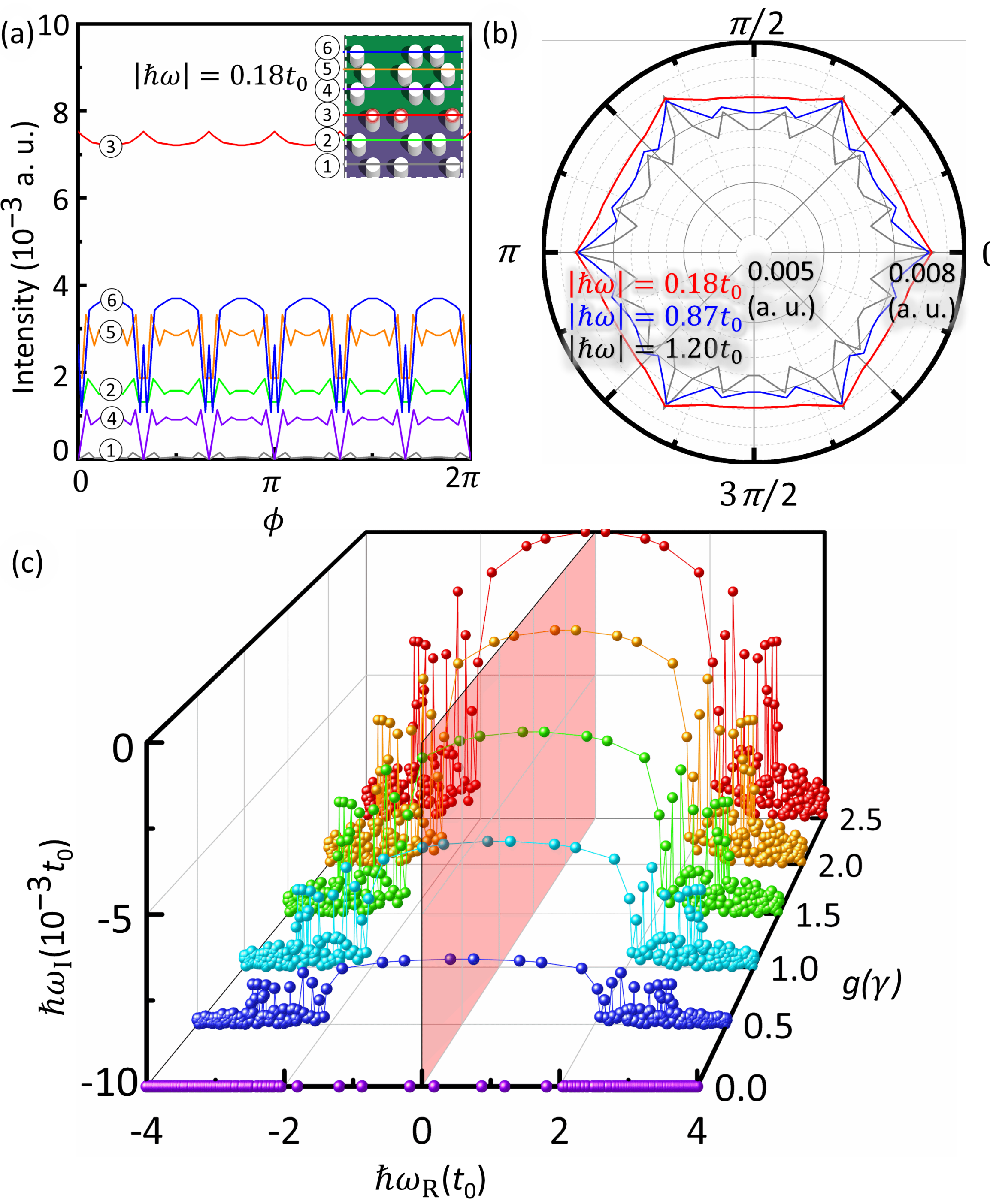

FIG. 5. (a) Distribution of intentity measured by the squared amplitude of the four states $|\pm 1/2, \pm\rangle$ along the radial direction. (b) Distributions of the intensity along the azimutal direction on the third layer (③ in (a)), where the maximal amplitudes are taken for the TRC modes as seen in (a). (c) Gain-value dependence of the complex eigenvalues. The imaginary eigenvalues of the four states $|\pm 1/2, \pm\rangle$ are always the largest among all eigenstates, and become positive first at threshold value $g_{\mathrm{th}} = 2.5\gamma$. Loss is set as $\gamma = 0.01t_0$ in the present work, which corresponds to constant complex eigenvalues at $g = 0$ shown by purple dots.

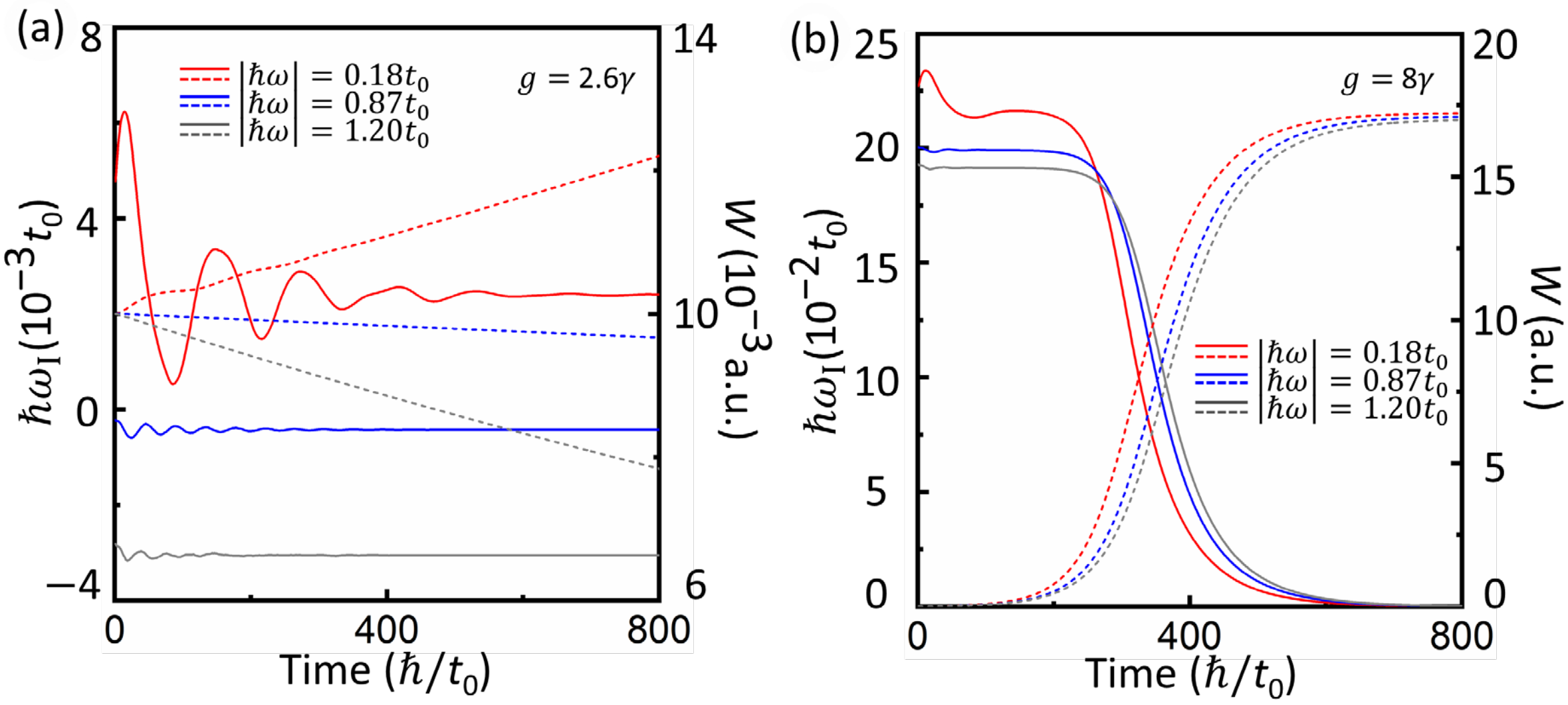

FIG. 6. (a), (b) Time evolution of the imaginary eigenvalue (solid curve) and wave function amplitude (dashed curves) for $g = 2.6\gamma$ and $g = 8\gamma$, respectively. All other modes take smaller imaginary eigenvalues than those shown explicitly.

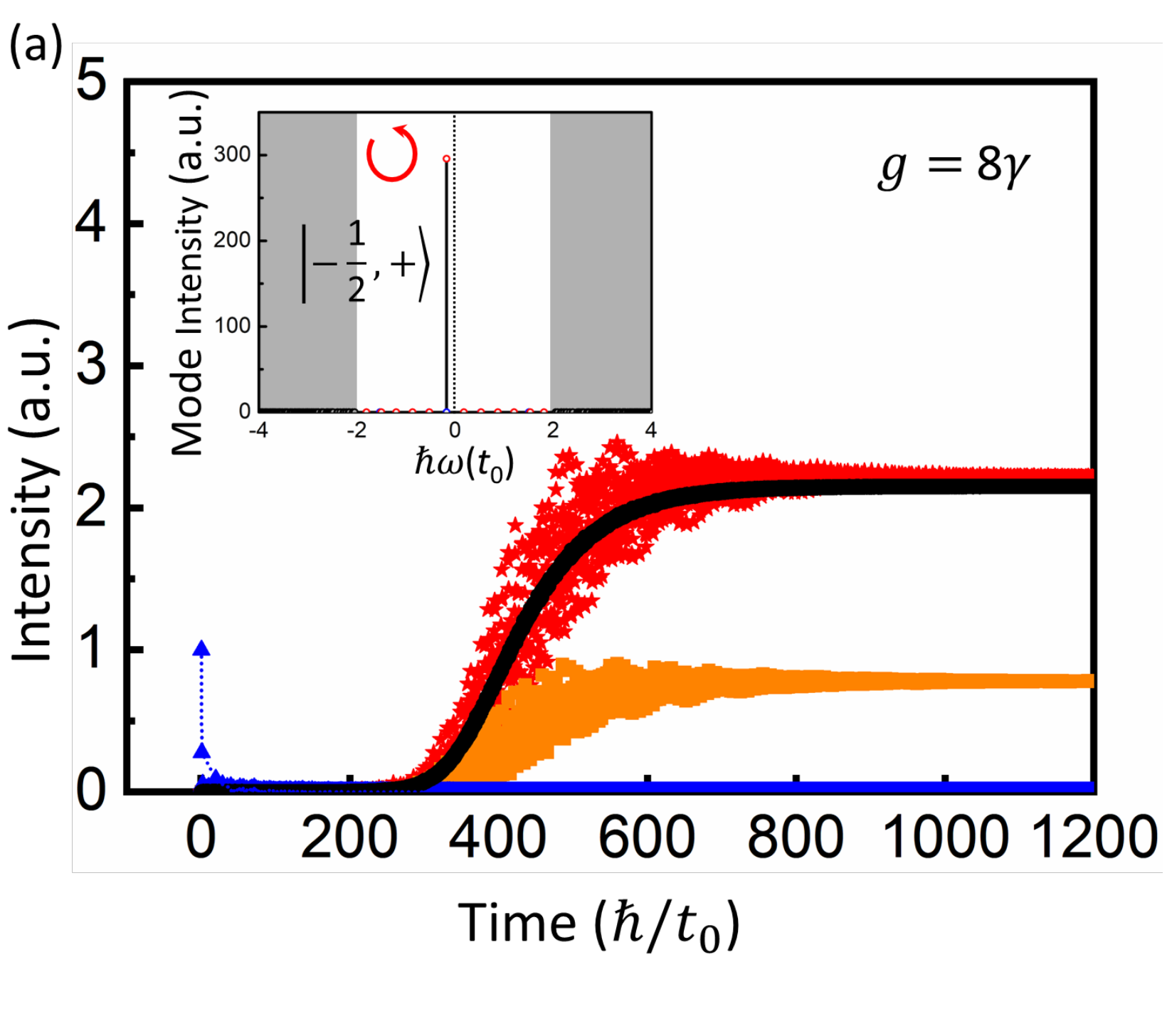

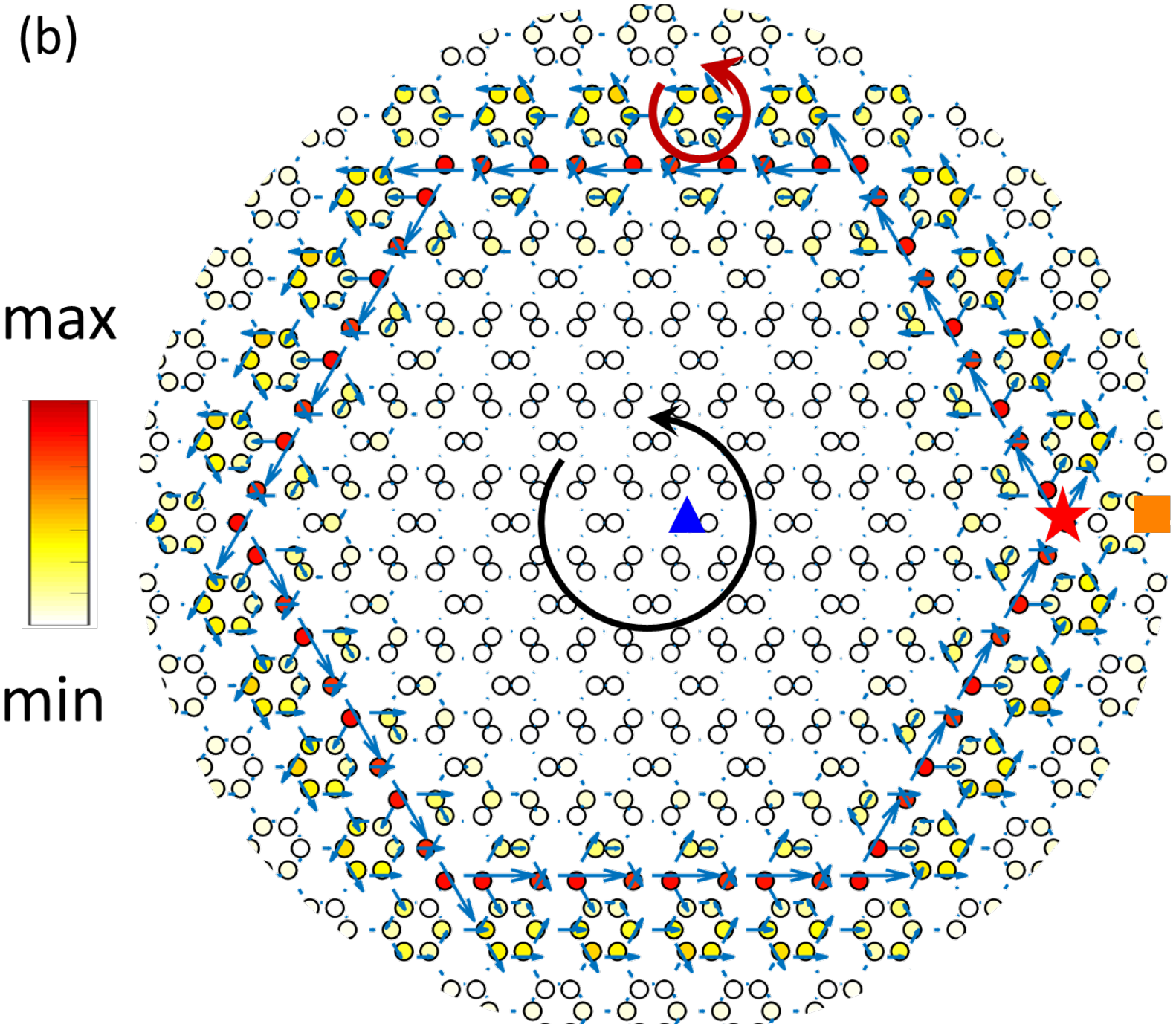

FIG. 7. (a) Time evolutions of intensity at the three sites indicated in (b) from the initial state located at the central unit cell with $2\pi$ phase winding, where the black solid curve represents the average value of intensities on sites on the interface. The inset is the power spectrum of the saturated state. (b) Distribution of the saturated wave function intensity and local energy flow given by Poynting vector (blue line arrows), whereas red (black) circular arrow shows the energy flow in a unit cell (along the ring cavity) as a guide of eyes.

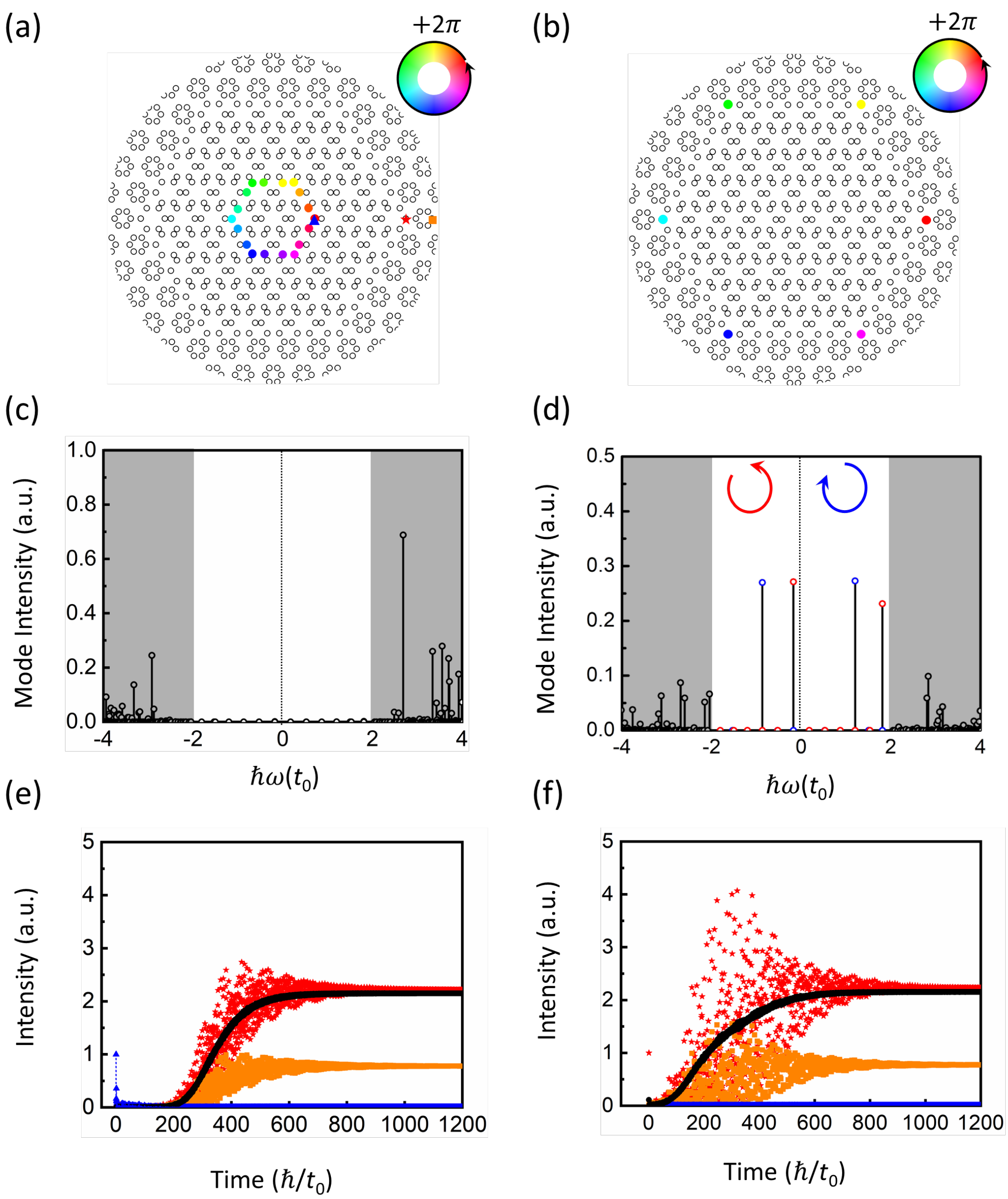

Fig 8. (a) Field distribution for the initial state with finite wavefunctions assigned at the device center, where colors represent phases. (b) Same as (a) except for that the initial state is at six corner sites in the ring cavity. (c), (d) Power spectrum for the initial state in (a) and (b), respectively. (e), (f) Time evolutions of intensity at the three sites indicated in (a) with (a) and (b) as the initial state, respectively, where the black solid curve represents the average intensity on interface sites.

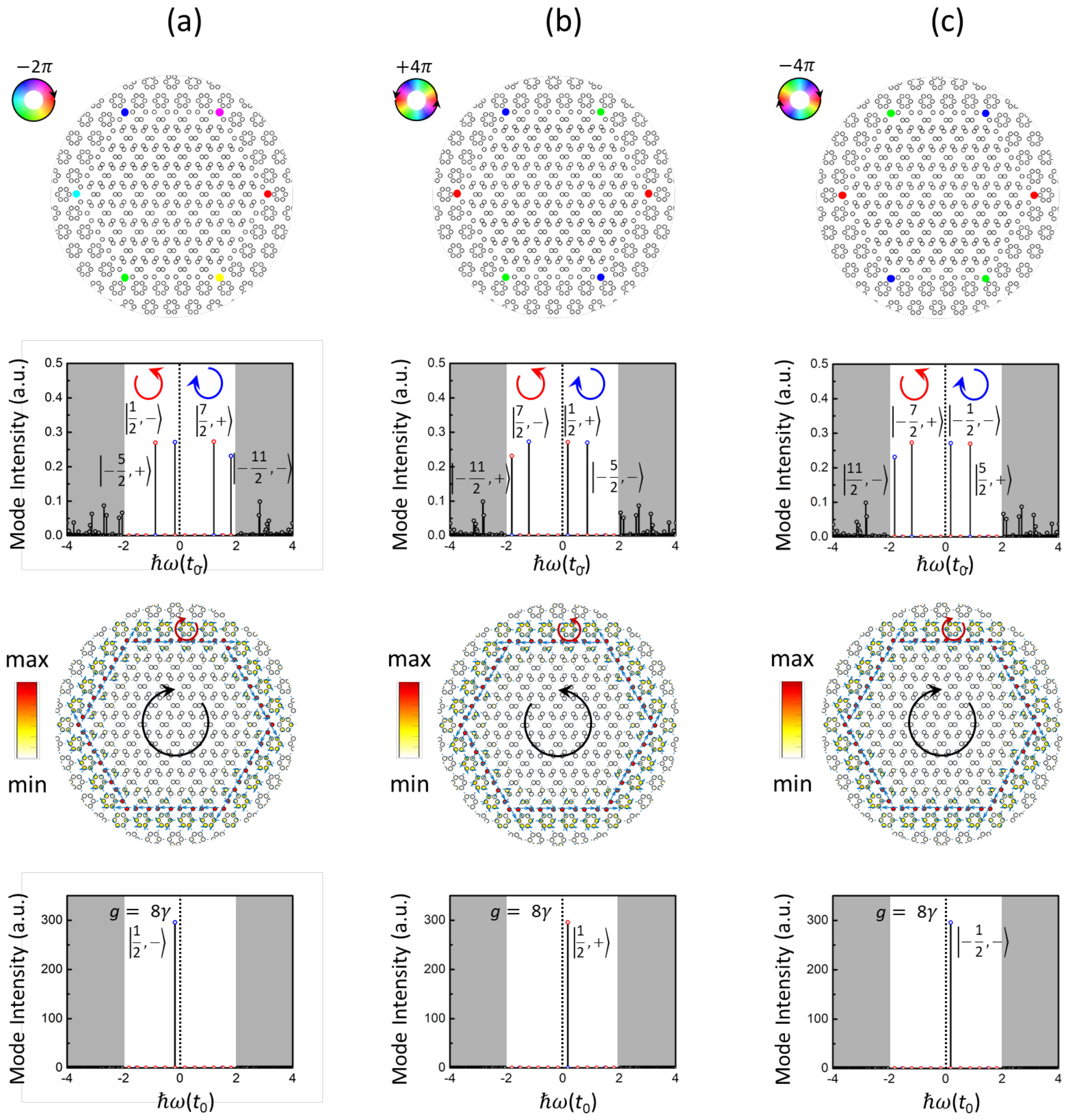

FIG. 9. Selective stimulation of lasing modes characterized by OAM. (a)-(c) Initial wave functions same as those shown in Fig. 8(b) except that the phase gradient is set as $-\pi/3$, $2\pi/3$ and $-2\pi/3$, and the lasing modes are $|1/2,-\rangle$ $|1/2,+\rangle$ and $|-1/2,-\rangle$, respectively.

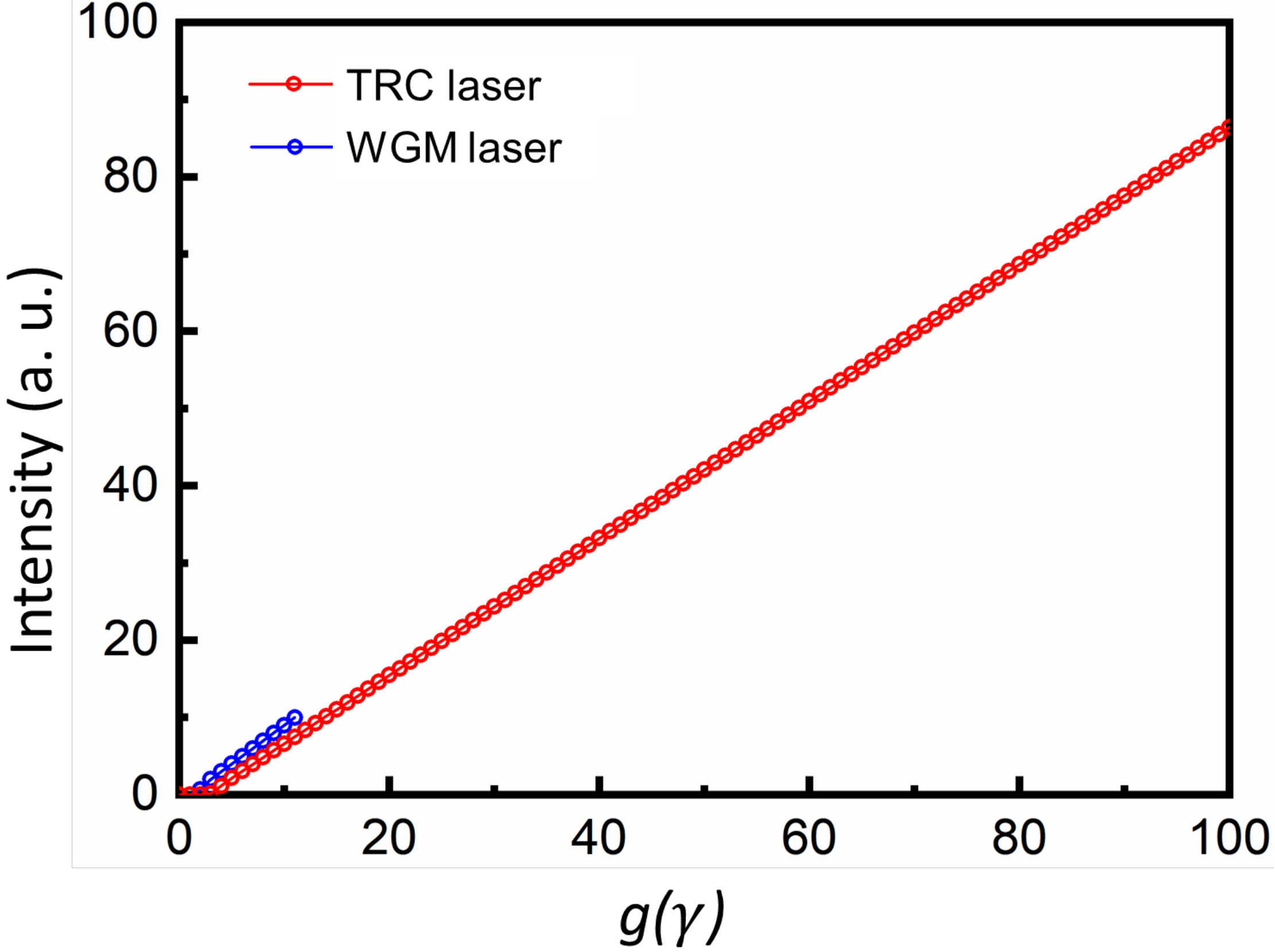

FIG. 10. Intensity-gain relation of the $|-1/2,+\rangle$ mode with the lasing threshold $g_{\mathrm{th}} = 2.5\gamma$. Parameters are the same as Fig. 7. As a comparison, the same quantity is displayed for a whispering-gallery-mode (WGM) laser with $g_{\mathrm{th}} = 1.0\gamma$. It is noticed that the single-mode lasing of the TRC modes is stable up to $g = 370\gamma$, a huge gain value much beyond the scale.

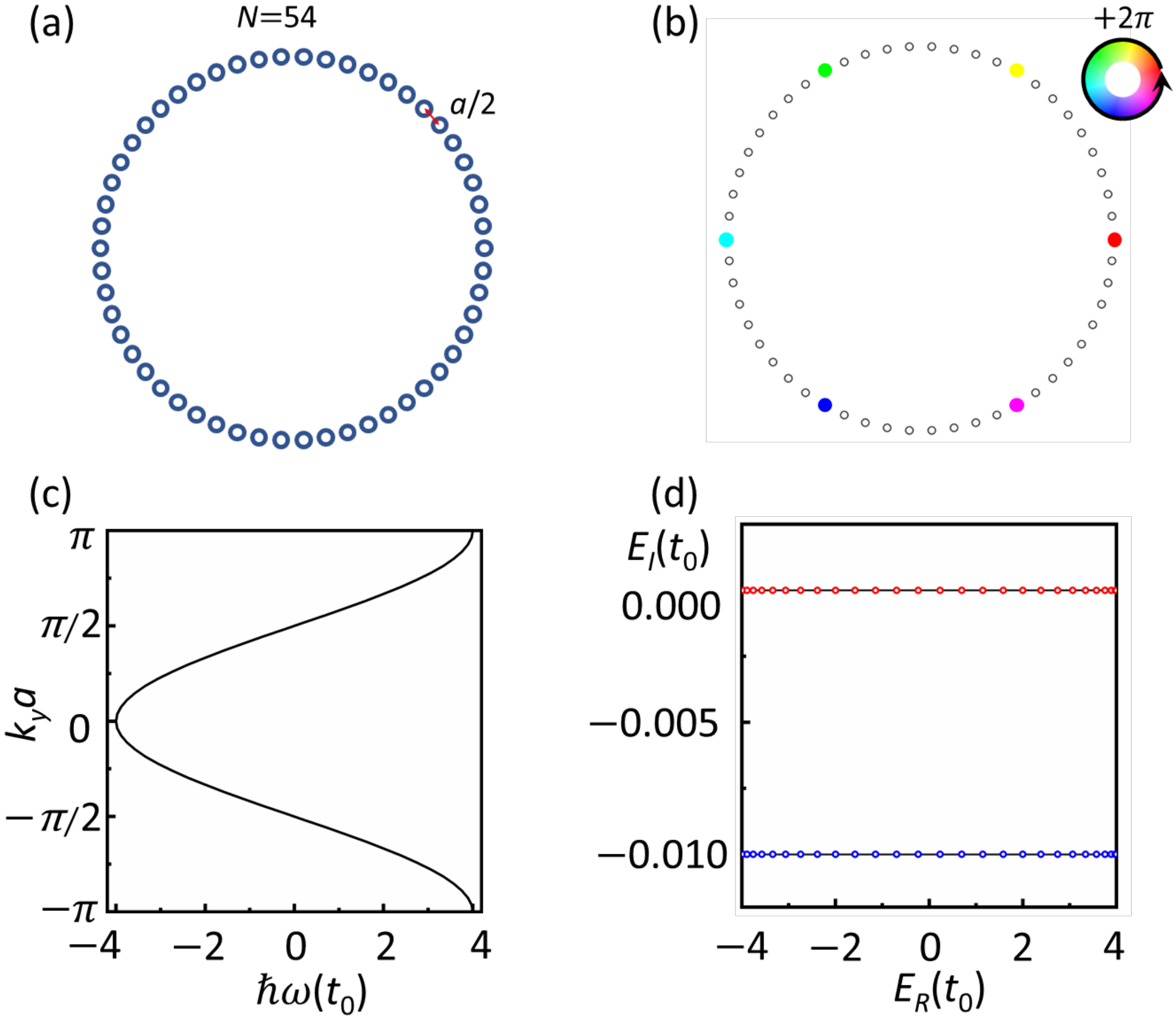

FIG. 11. (a) Schematic 1D circular chain for the WGM laser. (b) Initial state similar to those for the TRC laser. (c) Cosine dispersion relation of WGM. (d) Complex eigenvalues of the WGM laser at $g = 0$ and $g = \gamma (= g_{\mathrm{th}})$.

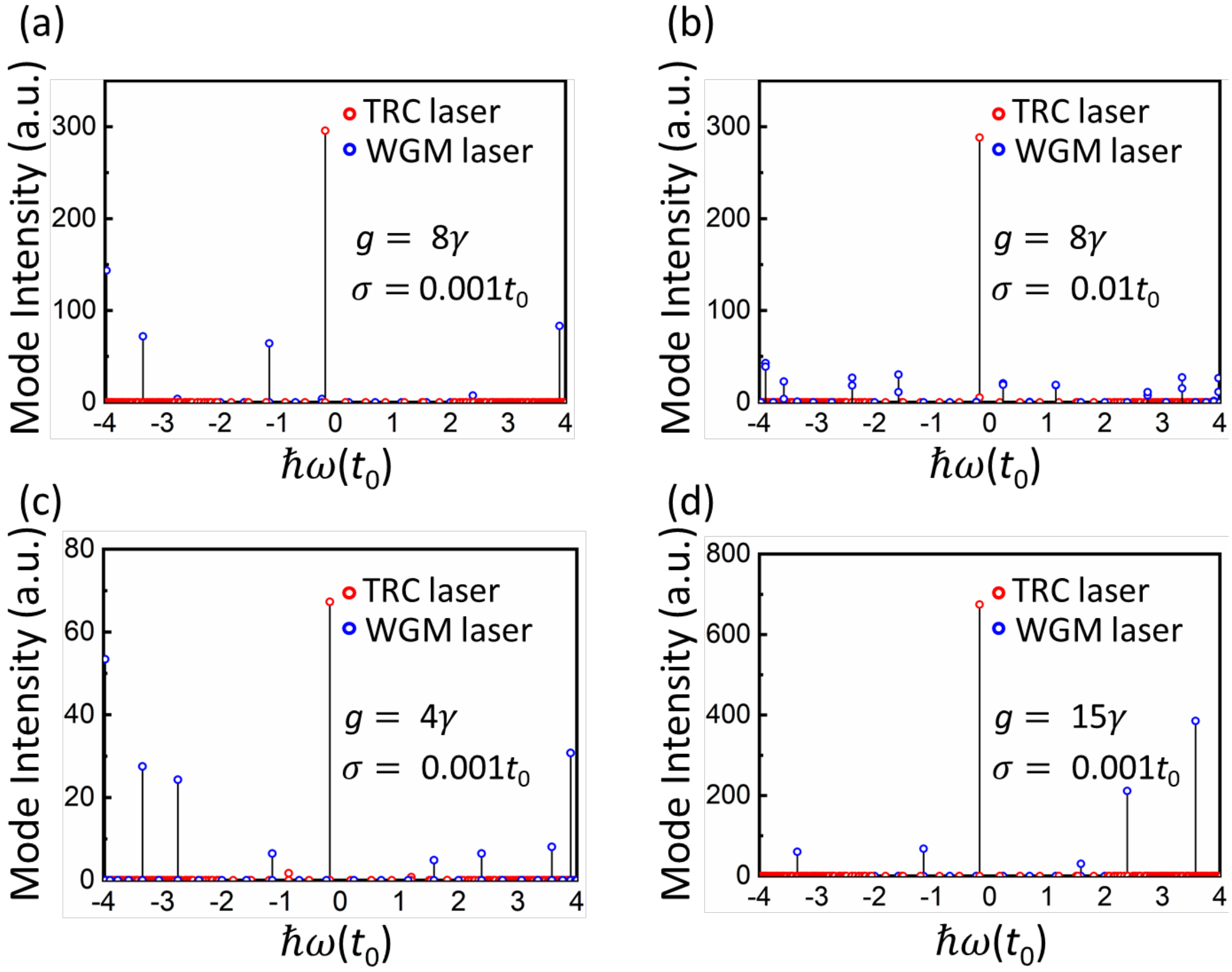

FIG. 12. Spectra for the WGM laser and TRC laser with random potentials. While the spectrum of the WGM laser is scattered by random potentials, the single-mode lasing remains intact in the TRC laser.

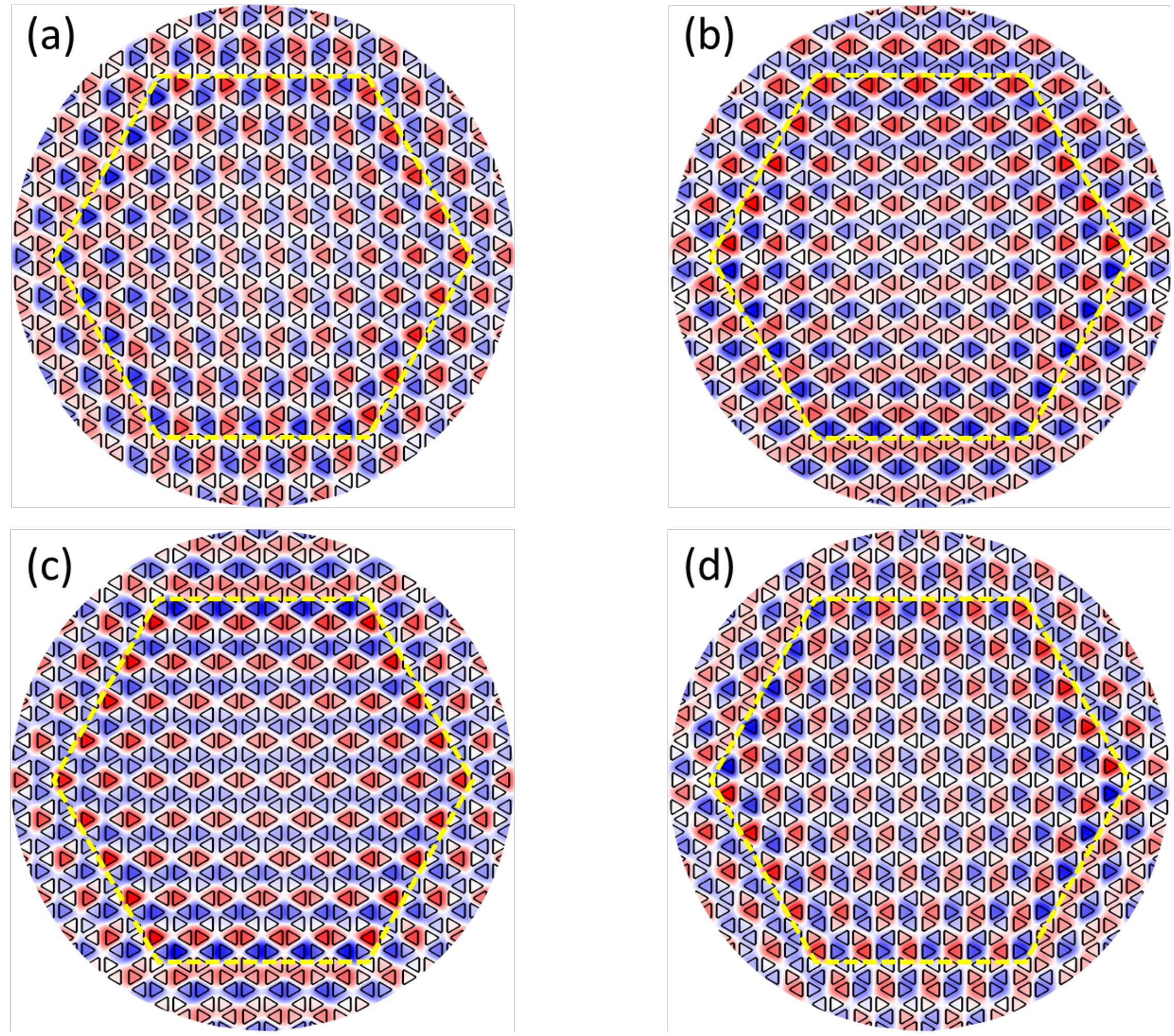

FIG. 13. Wave function of the ring-cavity modes obtained by COMSOL for the PhCs made of air-hole arrays in dielectric slabs. The honeycomb-type lattices are with parameters $a = 450\text{nm}$, $R_< = 1.05a/3$ for the topological PhC, $R_> = 0.94a/3$ for the trivial PhC and $\epsilon = 12.3\epsilon_0$ in the slab. The side length of the triangular hole is $0.2\sqrt{3}a$ and the vertex angles of the triangular hole are rounded to arcs with radius $r = 0.025a$. (a), (b) Real and imaginary parts of the TRC mode $|-1/2,+\rangle(= \text{Re} + i\text{Im})$ with a global OAM $L = +1$. (c), (d) Real and imaginary parts of the TRC mode $|1/2,+\rangle(= \text{Re} + i\text{Im})$ with a global OAM $L = +2$.